\documentclass[fleqn,usenatbib]{mnras}

\usepackage{newtxtext,newtxmath}

\usepackage[T1]{fontenc}
\usepackage{multirow}
\usepackage{subcaption}
\usepackage{import}
\usepackage{xcolor}
\usepackage{color}
\usepackage{adjustbox}
\usepackage{import}
\usepackage{lscape}

\DeclareRobustCommand{\VAN}[3]{#2}
\let\VANthebibliography\thebibliography
\def\thebibliography{\DeclareRobustCommand{\VAN}[3]{##3}\VANthebibliography}

\usepackage{graphicx}	
\usepackage{amsmath}	

\title[NEEDLE classifying rare transients]{NEural Engine for Discovering Luminous Events (NEEDLE): identifying rare transient candidates in real time from host galaxy images}

\author[Sheng et al.]{Xinyue Sheng,$^{1}$\thanks{E-mail: xsheng03@qub.ac.uk}
Matt Nicholl,$^{1}$
Ken W. Smith,$^{1}$
David R. Young,$^{1}$
Roy D. Williams,$^{2}$
\newauthor
Heloise F. Stevance,$^{3}$
Stephen J. Smartt,$^{3,1}$
Shubham Srivastav,$^{3,1}$
Thomas Moore$^{1}$
\\
$^{1}$Astrophysics Research Centre, School of Physics and Astronomy, Queen's University, Belfast, BT7 1NN, UK\\
$^{2}$ Institute for Astronomy, University of Edinburgh, Royal Observatory, Blackford Hill, EH9 3HJ, UK \\
$^{3}$ Department of Physics, University of Oxford, Denys Wilkinson Building, Keble Road, Oxford OX1 3RH, UK \\
}

\date{Accepted XXX. Received YYY; in original form ZZZ}

\pubyear{2023}

\begin{document}

\label{firstpage}
\maketitle

\begin{abstract}
Known for their efficiency in analyzing large data sets, machine learning-based classifiers have been widely used in wide-field sky survey pipelines. The upcoming Vera C. Rubin Observatory Legacy of Time and Space Survey (LSST) will generate millions of real-time alerts every night, enabling the discovery of large samples of rare events. Identifying such objects soon after explosion will be essential to study their evolution. 
This requires a machine learning framework that makes use of all the information available, including light curve, host galaxy and other contextual data.
Using $\sim5400$ transients from the ZTF Bright Transient Survey as training and testing data, we develop \textit{NEEDLE}, a novel hybrid (convolutional neural network $+$ dense neural network) classifier to select for two rare classes with strong environmental preferences: superluminous supernovae (SLSNe) preferring dwarf galaxies, and tidal disruption events (TDEs) occurring in the centres of nucleated galaxies. 
The input data includes (i) cutouts of the detection and reference images, (ii) photometric information contained directly in the alert packets, and (iii) host galaxy magnitudes from Pan-STARRS. Despite having only a few tens of examples of the rare classes, our average (best) completeness on an unseen test set reaches 77\% (93\%) for SLSNe and 72\% (87\%) for TDEs.
While very encouraging for completeness, this may still result in a large fraction of false positives (relatively low purity) for the rare transients, given the large class imbalance in real surveys. However, the goal of \textit{NEEDLE} is to find good candidates for spectroscopic classification, rather than to select pure photometric samples. Our network is designed with LSST in mind and we expect performance to improve further with the higher resolution images and more accurate transient and host photometry that will be available from Rubin. Our system will be deployed as an annotator on the UK alert broker, \textit{Lasair}, to provide predictions to the community in real time.
\end{abstract}



\section{Introduction}

Thanks to modern time-domain sky surveys, such as the Zwicky Transient Facility (ZTF; \citealt{Bellm_2019}), Asteroid Terrestrial impact Last Alert System (ATLAS; \citealt{Tonry_2018}), Panoramic Survey Telescope and Rapid Response System (Pan-STARRS;  \citealt{chambers2019panstarrs1}), and All-sky Automated Search for Supernovae (ASAS-SN; \citealt{Shappee_2014}), increasing numbers of transients have been discovered, catalogued and studied. The diversity of their spectra, and even photometric properties such as absolute magnitude, rise and decline rate and duration, have led to the identification of new and rare classes of events.
Even more encouraging is that the upcoming Legacy Survey of Space and Time (LSST; \citealt{Ivezic_2019}) survey will significantly increase the transient discovery rate through deeper observations, wide field of view and colour information from six filters.

In recent years, novel and rare superluminous supernovae (SLSNe) and tidal disruption events (TDEs) have been intensively studied, although their intrinsic physical mechanisms remain unclear.
SLSNe are around $\sim10$ times brighter than a Type Ia supernova (SN) and $\sim100$ times brighter than core-collapse SNe. Their rise timescales of greater than $\gtrsim 15-30$ days are also longer than typical supernovae \citep{Quimby_2011, Gal-Yam_2019, Nicholl_2021}. The hydrogen-poor events, also termed SLSNe Type I, mostly occur in low-mass dwarf galaxies with high specific star-formation rates and low metallicities \citep{Lunnan_2014,Leloudas_2015, Perley_2016, Angus_2016, Schulze_2017, Chen_2017}, which provides important hints for finding and identifying such events. 
Studying SLSNe allows researchers to fill gaps in our understanding of stellar evolution, particularly core-collapse supernovae in low-metallicity environments, and explore the extreme mass loss and rotation of possible massive progenitor stars.

A Tidal Disruption Event (TDE) occurs when a star's orbit gets close enough to be disrupted by the massive black hole (MBH) at the centre of a galaxy, leading to accretion onto the MBH with luminous emission and possibly jets \citep{Hills_1975, Rees_1988,Gezari_2021}.
Such rare events provide researchers with an opportunity to conveniently investigate accretion flows on quiescent black holes (at the low end of the MBH mass distribution), with accretion rates that change by orders of magnitude on human timescales.
Compared with SNe and SLSNe, TDEs can be differentiated by their locations in centres of their host galaxies as well as light curves that show a constant temperature. 
This provides helpful information for machine learning algorithms to learn their unique features.

Modern research on transients is mainly conducted on their spectra in the frequency domain and photometric information in the time domain. Spectra are essential to reveal their chemical compositions and physical properties (mass, velocity, redshift, etc). 
However, as photometric images require much shorter exposure time than spectra, they are preferred for observing missions that pursue large night sky coverage and long-term repeated detection.
As the number of transients discovered in wide-field imaging sky surveys grows exponentially, it is no longer possible to obtain spectra for most transients due to the expensive exposure times required.

The Vera C. Rubin Observatory (VRO) is planning to conduct LSST starting in 2025 \citep{Ivezic_2019}. 
LSST will observe the whole Southern sky and part of the Northern sky, including a Wide-Fast-Deep field ($90\%$ of the observing time) with seasonal cadence and a Deep-Drilling field with dense and deep detection.
Alert brokers, such as the UK alert broker \textit{Lasair} \citep{Smith_2019}, will provide researchers with real-time (within minutes to days) access to transient data. 
LSST is predicted to detect about 10 million transient alerts (defined as detections of time-varying flux) per night \citep{Kantor_2014}. These alerts will include $\sim10^4$ SLSNe \citep{Villar_2018} and $3,500-8,000$ TDEs \citep{Bricman_2020} per year. However, the number of conventional SNe detected each year will be $\gtrsim10^6$, meaning that only a small fraction of events will ever be observed spectroscopically. It is therefore essential to identify the most interesting candidates photometrically, in order to prioritise them for spectroscopy.

Machine learning algorithms will play an important role in classifying and filtering these alerts in real-time. This project aims to build up a hybrid classifier that fully takes advantage of various machine learning algorithms and combines different astronomical resources to identify candidate rare transients, such as SLSNe and TDEs, at or before their luminosity peak. For this reason, we are motivated to use only the properties available at the time of an early photometric detection: the early light curve, the associated discovery and reference images, and any cataloged host galaxy, but no information (such as redshift) that would require additional observations. We call this classifier the NEural Engine for Discovering Luminous Events (NEEDLE).

The paper outline is below: Section \ref{sec: state of the art} reviews some of the existing techniques in machine learning classification and why SLSNe and TDEs are promising targets. Section \ref{sec: data source} illustrates the data sources from ZTF bright transient survey and Pan-STARRS, and analyses the correlations between different features and transients. Section \ref{sec: preprocessing} describe the image and metadata pre-processing methods, including a binary classifier to assess image quality. Section \ref{sec: architecture} shows the model architecture, training and test sets, and development details. Section \ref{sec: results} shows the performance of the classifiers by confusion matrix as well as their completeness and purity diagrams, and illustrates the pipeline of \textit{NEEDLE} to provide classifications publicly on \textit{Lasair}. Then, Section \ref{sec:discussion} discusses the transient labelling issues, and comparisons with currently popular classifiers, and difficulties and improvements. Finally, Section \ref{sec: conclusion} is the summary of this paper.

\section{Contextual and machine learning classification}
\label{sec: state of the art}

\subsection{The host galaxy matters}
\label{sec: host}

The environment of transients have shown strong correlations with transients properties.
For example, the rates of typical Type Ia and core-collapse SNe scale with host galaxy stellar mass \citep{Sullivan_2006,Li_2011}. The relative fractions of different SN types vary between galaxies with different masses \citep{Graur_2017} and star-formation rates \citep{Botticella_2017}. 
The locations within their hosts vary, with some types of SNe also showing strong preferences for occurring in the brightest or bluest parts of their hosts \citep{Fruchter_2006,Kelly_2012, Blanchard_2016}. The rare transient classes that we are interested in, SLSNe and TDEs, are prime candidates for selection via their environments, as each shows strong biases in their host galaxies.

SLSNe are very unusual in that they show a strong preference (shared only by long gamma-ray bursts) for dwarf star-forming galaxies. SLSN samples also show a high fraction of irregular or interacting galaxies \citep{Chen_2017, Orum_2020}, but overall occur in low-density environments rather than groups or clusters \citep{Cleland_2023}. 
The locations of SLSNe within their hosts broadly track an exponential disk profile, but many events also occur at large offsets or in regions of low UV flux \citep{Hsu_2023}.

TDEs, like active galactic nuclei (AGN), occur at the centres of galaxies hosting MBHs. However, TDEs are rarely observed in galaxies with masses above $\sim{\rm few}\times 10^{10} {\rm M}_\odot$ \citep{van_Velzen_2021, Ramsden_2022}, since for very massive black holes the disruption occurs inside the event horizon. TDE hosts in particular show a large over-representation of recently quenched \citep{French_2016} galaxies with green colours \citep{Hammerstein_2022, Yao_2023}.
Compared to typical galaxies, their light profiles tend to be strongly peaked towards the nucleus \citep{Law-Smith_2017,Graur_2018}.

Some existing codes employ the context of where a transient appears to aid classification. For example, \textit{Sherlock}, applied on \textit{Lasair} is an integrated massive database system that classifies transients by cross-matching the position of a transient with all major astronomical catalogues
\citep{Smith_2020}. 
By associating transients with galaxies, galaxy nuclei, known AGN, variables or very bright stars, \textit{Sherlock} provides a top-level classification of any transient as a likely SN, nuclear transient, AGN, etc. 
Similarly, using contextual information, the \textit{ALeRCE} Stamp Classifier takes the first images and alert metadata for an object to provide a preliminary classification of AGN, SN, variable star, asteroid or bogus \citep{Carrasco_Davis_2021}.
\citet{Baldeschi_2020} presents a Random Forest (RF) classifier for galaxy classification based on recent star formation history and morphology, and applies it to the hosts of core-collapse and thermonuclear SNe, indicating that the colours and shapes of hosts can help the separation between two classes, better than random guessing.

Other codes go further and attempt to predict the spectroscopic sub-type of transient. \cite{Foley_2013, Kisley_2022} use purely host galaxy photometry to provide the probabilities of different types of SNe. \textit{GHOST} \citep{Gagliano_2021} employs a novel gradient ascent method to find the associated host galaxies, and based on the features of hosts and angular offset, they apply a RF to distinguish SLSNe, Type Ia SNe and core-collapse SNe. 
\citet{Gagliano_2023} takes host properties and light curves as inputs to classify SNe Ia, SNe II, and SNe Ib/c, and obtain increasing accuracy with later phases.
In summary, different transients have unique preferences for where they occur, and these can help reveal their likely nature.

\begin{figure*}
    \centering
    \includegraphics[width=\textwidth]{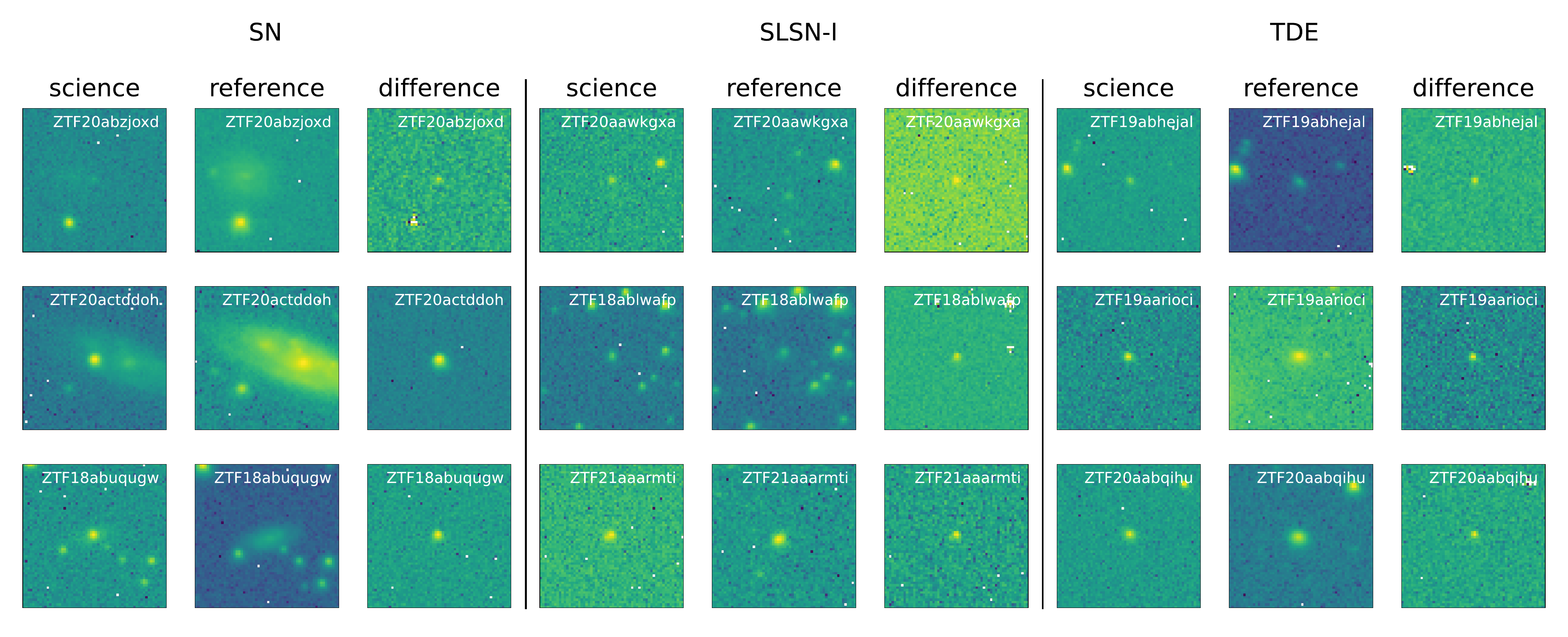}
    \caption{Examples of the \textit{science}, \textit{reference} and \textit{difference} images of three general classes: SLSN, TDE and SN.}
    \label{fig:three_classes}
\end{figure*}

\subsection{Machine learning architectures on transient classification}

Machine learning has been widely applied to astrophysical transients for classification tasks, such as BDT (SNGuess, \citealt{Miranda_2022}; Avocado, \citealt{Boone_2019}; \textit{Sherlock}, \citealt{Smith_2020}), RF (FLEET, \citealt{Gomez_2020}; \citealt{Baldeschi_2020}; ALeRCE light curve classifier, \citealt{Sanchez-Saez_2021}), and Neural Networks.
Particularly, deep learning algorithms (deep neural networks) have shown powerful performance in extracting features of data to improve classification without manual feature selection.

Recurrent Neural Networks (RNN) are capable of learning the correlations among close and distant time steps among time-series and are designed for classification, modelling and prediction. RNNs can extract features from light curves of different classes of transients to distinguish them. Examples of such codes include RAPID \citep{Muthukrishna_2019}, SuperRAENN \citep{Villar_2020}, Superphot \citep{Hosseinzadeh_2020}, Classifier for GOTO \citep{Burhanudin_2021} and Early-time transient Classifier \citep{Gagliano_2023}. 
Attention mechanism has also been applied, such as TimeModAttn \citep{Pimentel_2022}. 

On the other hand, Convolutional Neural Networks (CNN) are mainly designed for visual imagery classification. They can generate feature maps of the input data while training, and attempt to associate these features with class labels. Image-based classifiers have not yet attained the widespread use of light-curve classifiers, but experiments to date have shown that this approach is a very promising alternative, as it can take into account the transient position and host galaxy morphology, as discussed in \ref{sec: host}.
Transients researchers have been implementing CNNs for transient classification with codes such as ALeRCE \citep{Carrasco_Davis_2021}, DELIGHT \citep{Forster_2022}, and recent work on light curves by \citet{Burhanudin_2023}, proving that CNNs are able to achieve high accuracy in identifying various types of transients.

The above architectures have also shown promising performance in classifying rare events like SLSNe and TDEs.  
For SLSNe, classifiers using light curves achieve completeness $\sim0.69-0.83$, and in one case up to 1.00 completeness \citep{Qu_2022,Muthukrishna_2019,Sanchez-Saez_2021}. For TDEs, existing codes achieve 0.40 completeness \citep{Gomez_2023} at early phases, or better than 0.80 with full light curves \citep{Stein_2023}. A detailed review is shown in Table \ref{tab: comparison with classifiers}.

However, there are still difficulties and limitations remaining. 
CNNs may struggle with low image quality due to issues of signal-to-noise, resolution, bright nearby objects or detector cosmetics, resulting in mislabelling. 
Light curves with sparse cadences and few observations are difficult for RNNs to extract the correlations. 
For training and test data sets, the number of spectroscopically-confirmed SLSNe and TDEs make up only 1-2\% of all transient samples, leading to possible underrepresented learning.
Many classifiers, such as \citet{Hlovzek_2023}, have been trained on simulated data sets (e.g.~PlasTiCC; \citealt{Kessler_2019}), which avoids the difficulties that must be overcome when dealing with real data. 
Finally, any classifiers that require redshift information or the declining part of a light curve may not be suitable for early-time classification.

Novel architectures are required to gain better accuracy. In recent years, hybrid neural networks have become more popular. CNNs with artificial neural networks (ANN, fully connected neural networks) are able to fully use images and metadata (position, redshift, etc) together to provide high accuracy predictions (e.g. GaZNets, \citealt{Li_2022}; ALeRCE, \citealt{Carrasco_Davis_2021}).
Other architectures, such as transformers like ASTROMER \citep{Donoso-Oliva_2022}, use an Autoencoder with positional embedding and self-attention blocks to gain the representation of transients' light curves, which can be further applied to classification and modelling.

In short, more deep learning applications for astronomical study are expected to digest multivariate data. This might include magnitudes or fluxes in the time dimension, images (in one or more filters), and contextual information from existing catalogues. 
Our goal here is to take the first steps in realising a hybrid classifier that tries to maximise the information used from images, simple light curve features and host galaxy features, and apply this to the case of finding SLSNe and TDEs in wide-field surveys.

\section{Data Set} \label{sec: data source}

For this project we require a training and test set of transients with known classes (based on spectroscopic classifications). In this section we outline the sources of data used to train and validate our code.

\subsection{ZTF bright transients database}

Although our ultimate goal is to develop a classifier for LSST, for our initial training and test set before that survey begins we use the Zwicky Transient Facility (ZTF) Bright Transient Survey (BTS) \citep{Bellm_2019, Fremling_2020, Perley_2020}. The ZTF public survey covers the entire Northern sky to a depth of $\approx 20-20.5$\,mag every 2-3 nights in $g$ and $r$ filters. The BTS has been spectroscopically classifying all ZTF-detected supernovae brighter than $\approx 19$\,mag since June of 2018.
We choose this dataset as it is the largest homogeneous set of labelled transients available, and the data are comparable to LSST in terms of imaging cadence and the format of the real-time alerts.

We downloaded the entire ZTF BTS sample brighter than 19\,mag, up to March 2022, and use this as the basis of our sample.
This contains 5703 spectroscopically classified transients. 
Information, such as ZTF object ID, coordinates, discovery date and spectroscopic type can be found from ZTF Bright Transient Survey Sample Explorer\footnote{\url{https://sites.astro.caltech.edu/ztf/bts/explorer.php?f=s&subsample=trans&classstring=&endpeakmag=19.0&purity=y&quality=y}}.
After removing duplicates and missing objects in the ZTF database, 5388 ZTF objects are obtained.
This includes over 5000 SNe, but only 37 SLSNe and 18 TDEs. We therefore supplement the BTS data set with any SLSNe or TDEs published in ZTF sample papers. This includes TDEs from \citet{Hammerstein_2022,van_Velzen_2021} and SLSNe from \citet{Chen_2022}. Given that some of these objects are already in the BTS data, our total numbers of SLSNe and TDEs are 87 and 64, respectively.

All transients fall into five general categories, shown in Table \ref{tab: all_objects}: 
\begin{itemize}
    \item ``Common'' Supernovae (including all spectroscopic Type Ia, core-collapse, and interacting SNe)
    \item Superluminous supernovae (considering here only the hydrogen-poor SLSNe Type I)
    \item Tidal Disruption Events
    \item Possible SNe/transients of ambiguous nature (calcium-rich, gap transients)
    \item Non-SN (novae and stellar outbursts).
\end{itemize}

The latter two categories make up only ~1\% of the sample, and can generally be filtered out by their fast light curves before being passed to the machine learning classifier. We do not include them in our training or test set, but include them in the Appendix for completeness.
The first category is very broad, containing 97.2\% of events.
However, as the task of \textit{NEEDLE} is to distinguish among SNe, TDEs and SLSNe, we avoid sub-dividing the SN class so that more attention can be focused on the rare classes of interest. This is the first version of NEEDLE, our aim is that future versions with improved architecture and more data will be able to perform more fine-grained classification of the various supernova sub-types.
Table \ref{tab: nums of classes} provides counts of objects with image and magnitude data from ZTF, as well as the numbers that also have cataloged host data in $g$ and $r$ bands from deeper surveys.

\begin{table}
\begin{tabular}{c|ccc|ccc}
\hline
Band          & \multicolumn{3}{c|}{g}                                        & \multicolumn{3}{c}{r}                                         \\ \hline
Label         & \multicolumn{1}{c|}{SN}   & \multicolumn{1}{c|}{SLSN} & TDE & \multicolumn{1}{c|}{SN}   & \multicolumn{1}{c|}{SLSN} & TDE \\ \hline
Object        & \multicolumn{1}{c|}{5185} & \multicolumn{1}{c|}{80}     & 62  & \multicolumn{1}{c|}{5237} & \multicolumn{1}{c|}{87}     & 64  \\ \hline
Object \& Host & \multicolumn{1}{c|}{4959} & \multicolumn{1}{c|}{37}     & 60  & \multicolumn{1}{c|}{5016} & \multicolumn{1}{c|}{41}     & 62  \\ \hline
\end{tabular}
\caption{The number of objects in each class with images and light curve information from ZTF, as well as those with host galaxy matches in existing catalogs from \textit{Sherlock}. The information is provided separately for the $g$ and $r$ bands. }
\label{tab: nums of classes}
\end{table}

\begin{figure*}
    \centering
    \includegraphics[width=\textwidth]{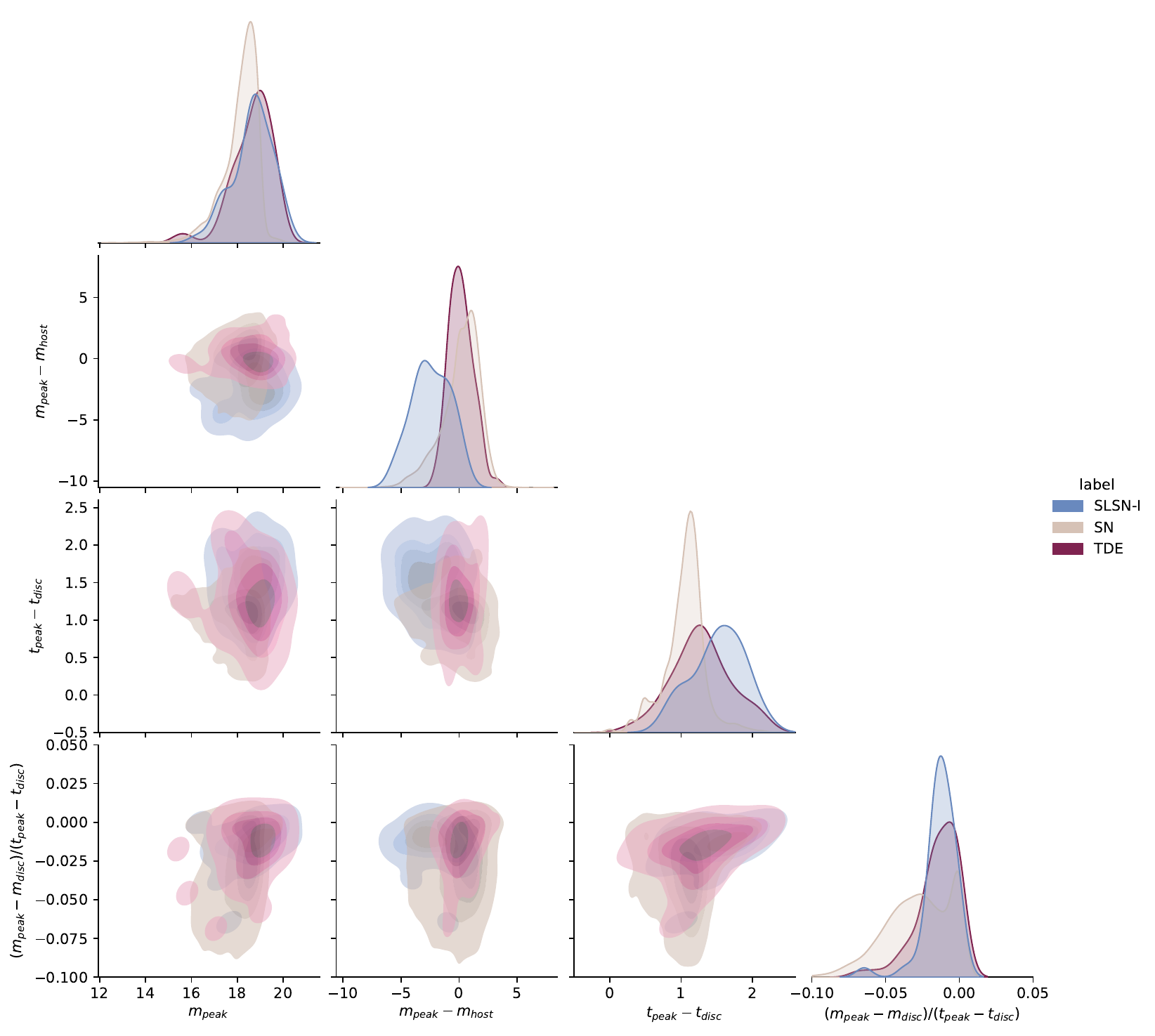}
    \caption{Corner plot showing distributions of selected light curve features in the data set. To show the difference between distributions more clearly, the light curve slope $(m_{\rm peak}-m_{\rm disc}/(t_{\rm peak}-t_{\rm disc})$ and rise time $t_{\rm peak}-t_{\rm disc}$ 
    have been scaled logarithmically.
    }
    \label{fig: corner plot transient}
\end{figure*}

\begin{figure*}
    \centering
    \includegraphics[width=\textwidth]{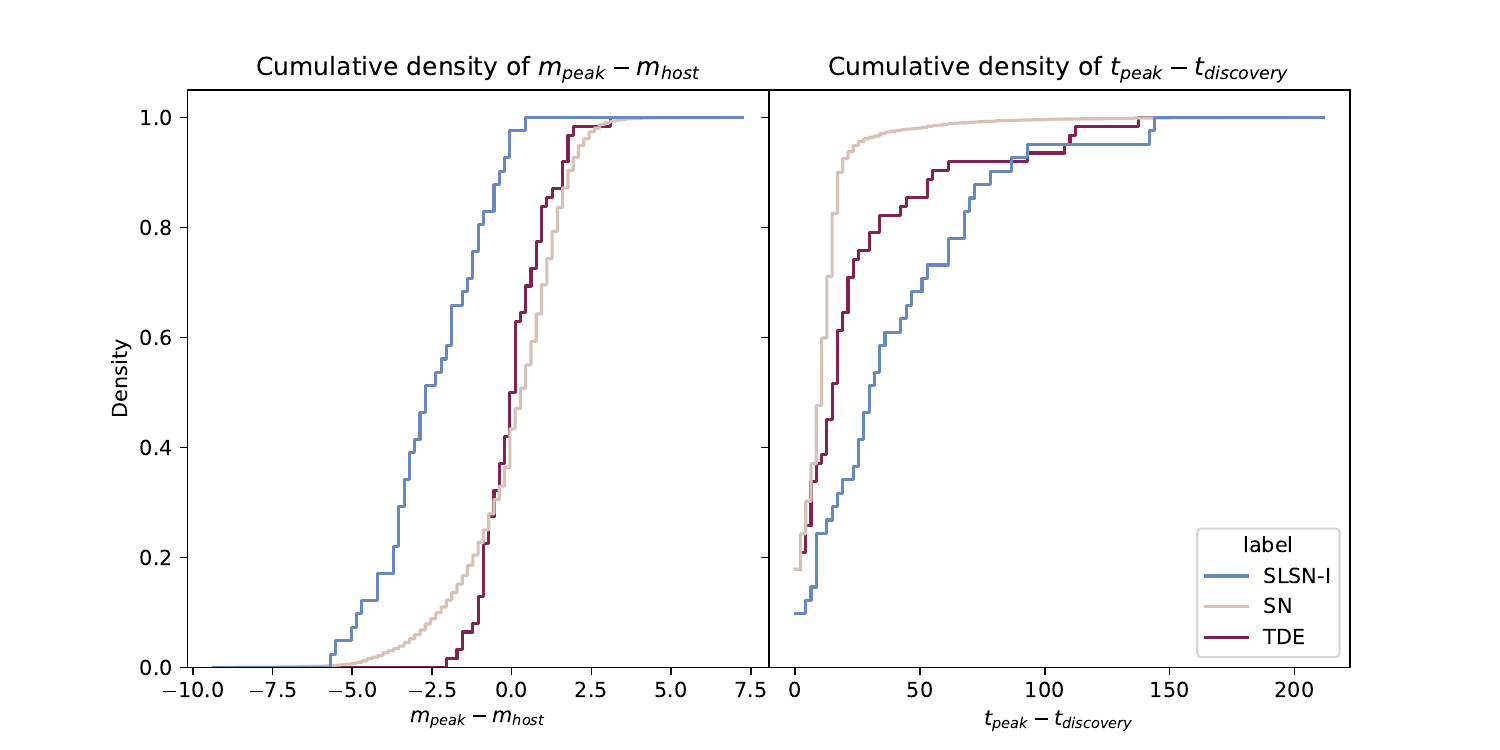}
    \caption{Cumulative distribution functions for transient contrast with host galaxy $m_{\rm peak}-m_{\rm host}$ and approximate rise time $t_{\rm peak}-t_{\rm discovery}$.
    }
    \label{fig: cumulative plots}
\end{figure*}

\begin{figure*}
    \centering
    \includegraphics[width=\textwidth]{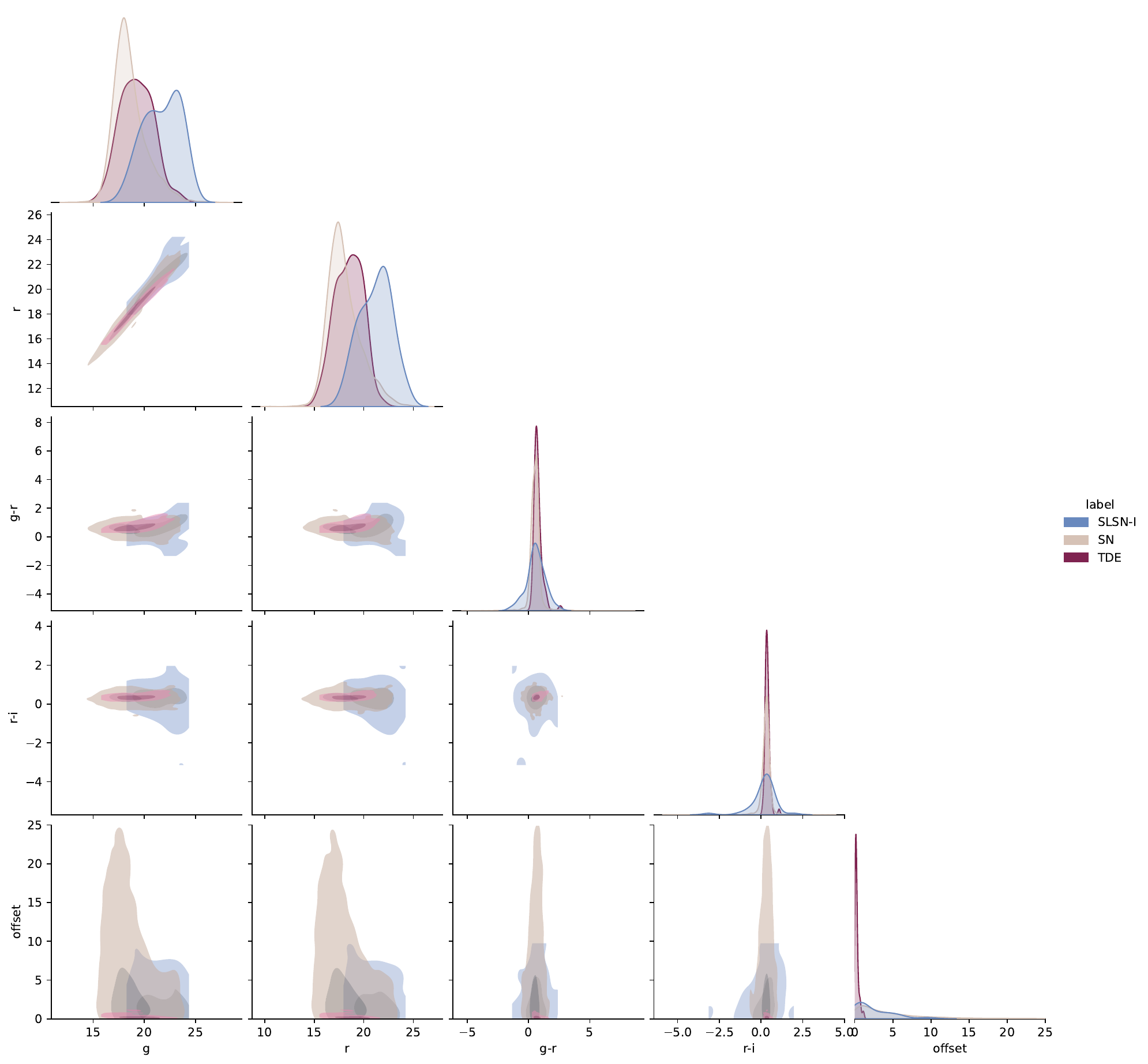}
    \caption{Corner plot showing host galaxy magnitudes in $g$, $r$ and $i$ bands, as well as the colours in $g-r$ and $r-i$. We also show the \textit{offset}, meaning the projected distance between the transient and its likely host in arcseconds.
    }
    \label{fig: corner plot host}
\end{figure*}

\subsection{Images}

We wrote Python scripts to download ZTF cutout images centered at the transient positions, using the ZTF image database API. 
For each ZTF object, starting from its discovery date to 200 days later, we downloaded all available images in the $g$ and $r$ bands. This includes: the \textit{Science} image - the image taken in each visit, containing the transient flux; the \textit{Reference} image - a stacked image from before the event's discovery, providing a template of the host galaxy and surrounding field; and the \textit{Difference} image - the subtraction of the above two images, containing only transient flux. The requested image size is 1 arcmin. This size is large enough to include most host galaxies, while larger images would include more unrelated sources in the field. Figure \ref{fig:three_classes} shows examples of ZTF images obtained for three classes of transients we consider.

\subsection{Image metadata}

We label each image with metadata including ZTF object ID, class label (SN, SLSN or TDE), RA, Dec, image size, and date. We retrieve separately for each filter the start and end Julian dates where the object is detected. This information is stored in a JSON file for each object. Although we download all images and their associated metadata for these objects, we find better performance in training when we give our network only one image per object, therefore when training the model we use the image metadata to select one image from close to the time of light curve peak.

\subsection{Light curve metadata}

For each object, we also retrieve its photometry for each available detection through \textit{Lasair}. Our aim is to include some simple light curve parameters (features) as additional data for our classifier. We use the \textit{Lasair} API to query the light curve using the ZTF object ID.
After cross-matching with the image data, we found that not every image has a corresponding magnitude, as \textit{Lasair} contains only the public ZTF photometry. Although more light curve data would be available by querying the ZTF forced photometry, using the data from \textit{Lasair} ensures consistent formatting between our training data set and future real-time alert classifications we wish to perform with our trained model. This light curve data (detection dates and magnitudes) is appended to the metadata file for each object. The simple features extracted are listed in Table \ref{tab: metadata}.

\subsection{Host galaxy metadata}

The \textit{Sherlock} software package  \citep{Smith_2020} is integrated into \textit{Lasair}, and automatically provides a contextual classification by cross-matching with a library of historical and on-going sky survey catalogs. This provides preliminary classifications as transients, variables, artefacts, etc, based on association with nearby galaxies, known cataclysmic variable stars, active galactic nuclei, or bright stars. For this project, we query the \textit{Sherlock} table on \textit{Lasair} to find the coordinates of the most likely host galaxy for each of our transients.

We use these coordinates to retrieve host galaxy magnitudes from Data Release 2 of the Pan-STARRS survey. 
Pan-STARRS is a wide-field imaging system that observes 30,000\,deg$^2$ of the Northern sky in five broadband filters ($g$, $r$, $i$, $z$ and $y$). The stacked depth is up to 23.3 mag, 23.2 mag, 23.1 mag, 22.3 mag, 21.3 mag, respectively \citep{chambers2019panstarrs1, Flewelling_2020}.
The Pan-STARRS footprint completely overlaps the ZTF coverage, and DR2 contains data taken between 2010 and 2014, prior to the ZTF survey.
Therefore, Pan-STARRS should contain all ZTF transient host galaxies brighter than $\sim 23$ mag.

Possible hosts were found by \textit{Sherlock} for most transients, but about half of the SLSN hosts are missed, as they are likely fainter than the Pan-STARRS DR2 limiting magnitude. This is unsurprising given that most SLSNe explode in distant dwarf galaxies. 
We used the Pan-STARRS DR2 API to obtain the Aperture magnitude in $g$, $r$, $i$, $z$ and $y$ bands. The colour of the host can be measured by $g-r$, and $r-i$, and is correlated with the age and star-formation rate of the stellar population.

The full list of metadata used in this study is given in Table \ref{tab: metadata}.

\begin{table*}
\begin{tabular}{|l|l|l|}
\hline
Metadata                            & Feature                & Definition \\ \hline
\multirow{6}{*}{Transients alerts} & $m_{peak}$             & Peak magnitude among all existing                                        observations. \\ \cline{2-3} 
                                    & $m_{discovery}$             & Magnitude of the discovery observation.\\ \cline{2-3} 
                                    
                                    & $\Delta T_{discovery}$      & Time difference between current observation and discovery observation            \\ \cline{2-3} 
                                    & $\Delta m_{discovery}$      & Magnitude difference between current observation and discovery observation\\ \cline{2-3} 
                                    & $\frac{\Delta m_{discovery}}{\Delta T_{discovery}}$ & Ratio of the magnitude difference over time difference since discovery\\ \cline{2-3} 
                                    & $\frac{\Delta m_{recent}}{\Delta T_{recent}}$ & Ratio of the magnitude difference over time difference since last detection\\  \hline
                                   
\multirow{9}{*}{Host galaxy}        & $m_{gAp}$              & Host magnitude in $g$ band \\                                         \cline{2-3} 
                                    & $m_{rAp}$              & Host magnitude in $r$ band \\ \cline{2-3} 
                                    & $m_{iAp}$              & Host magnitude in $i$ band \\ \cline{2-3} 
                                    & $m_{zAp}$              & Host magnitude in $z$ band \\ \cline{2-3} 
                                    & $m_{yAp}$              & Host magnitude in $y$ band \\ \cline{2-3} 
                                    & $m_{g-r}$              & Host magnitude difference between $g$ and $r$ bands           \\ \cline{2-3} 
                                    & $m_{r-i}$              & Host magnitude difference between $r$ and $i$ bands           \\ \cline{2-3} 
                                    & separationArcsec             & The distance between the host centre and transient on the image \\ \cline{2-3} 
                                    & $\Delta m_{host-peak}$ & Magnitude difference between the host magnitude and $M_{peak}$ \\  \hline

\end{tabular}
\caption{Summary of light curve and host galaxy features included in our metadata.}
\label{tab: metadata}
\end{table*}

\subsection{Data Analysis}\label{sec: data analysis}

Before building a model, we check for correlations or clusters within our metadata.
Figure \ref{fig: corner plot transient} shows simple features obtained from the ZTF $r$-band light curves. We show the apparent magnitude around the peak $m_{\rm peak}$, the magnitude contrast between transient and host, the elapsed time between first detection and light curve peak, as well as a measure of the light curve slope during the rise, the ``rising ratio'' ($m_{\rm peak}$ - $m_{\rm discovery}$)/($t_{\rm peak}-t_{\rm discovery}$). 

It can be seen that the distributions of SLSN and TDE apparent magnitudes in our sample skew dimmer than the distribution for other SNe. This is due in part to the lack of nearby events in these rare classes. The need to include examples from outside of the magnitude-limiting BTS sample may also bias these events towards fainter magnitudes, but is unavoidable given the class size imbalance. 

SLSNe show a much larger contrast at peak with their host galaxies, standing out from SNe and TDEs. 
Moreover, their rising timescales are longer than other transients. Normal SNe show the fastest rise, with TDEs showing a broad distribution peaking in between the other classes. The median SLSN rising ratio is similar to TDEs, but the deviation is smaller. To compare some of the key parameters more clearly, Figure \ref{fig: cumulative plots} presents the cumulative distributions of host galaxy contrast ($m_{\rm peak}$ - $m_{\rm host}$) and approximate rise time ($t_{\rm peak}$ - $t_{\rm discovery}$), where the three classes show clear differences.

Similarly, Figure \ref{fig: corner plot host} shows a corner plot for host galaxy metadata, including magnitudes and colours in $g$, $r$ and $i$ bands, and the offset in arcseconds between the transient coordinates and the host galaxy centroid. Again the plot shows that SLSNe tend to have the faintest hosts, with a slight bias to bluer colours in $g-r$. As expected, the host offset for most TDEs is clustered around $\sim0.0-1.0$ arcseconds. SLSNe, with their compact hosts, tend to show small offsets of a few arcseconds, whereas the distribution is much broader for typical SNe in extended galaxies.

\section{Data Preprocessing}\label{sec: preprocessing}

In this section, we illustrate the steps used to clean and prepare our data set before training our model.

\subsection{Image Preprocessing}  

\begin{figure*}
    \centering
    \includegraphics[width=\textwidth]{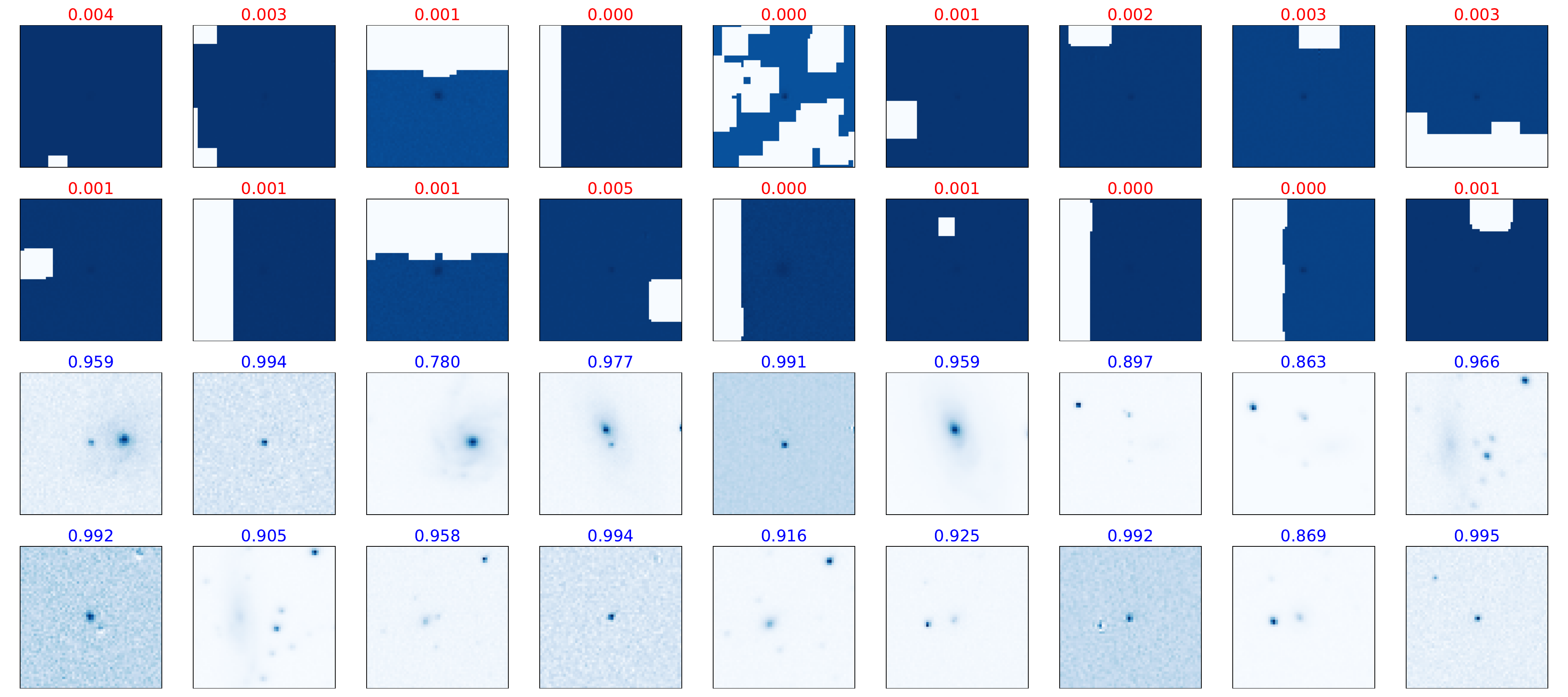}
    \caption{Bad and good images samples with the quality check classifier predictions. Images with red probabilities are judged as `bad images' and removed before feeding into \textit{NEEDLE}.
    }
    \label{fig: bad image samples}
\end{figure*}

Some of the images we obtained from the ZTF database were found to have irregular sizes, shapes and missing pixels. Such issues can be caused by a transient position close to the edge of the detector field of view, or nearby bright stars that are masked out (but can still leave diffraction spikes or subtraction artefacts). Examples are shown in Figure \ref{fig: bad image samples}. These poor quality images can severely impact the training process. Therefore, there are a range of ways to identify, modify or delete them before training.

\subsubsection{Image size cutout}

An image with a shape slightly smaller than 60x60 pixels (for example, 58x58 pixels) will be expanded to 60x60 pixels by repeating the last row or column on each side. However, for those with very small sizes, they will be removed. On the other hand, those larger than 60x60 pixels are reduced to 60x60 size.

\subsubsection{Quality check model}\label{sec: quality check}

Images with missing or unreliable pixels are tricky to deal with, and those bad images greatly harm the training process.
One common feature is that such images often have very large standard deviations ($\sigma$), much larger than normal images. 
However, our experiments showed that a quality cut based only on $\sigma$ still cannot get rid of a small number of problematic images that have reasonable standard deviations. 
Therefore, a binary convolutional neural network is developed to determine whether an image is good or bad.

\begin{figure*}
     \centering
     \begin{subfigure}[b]{0.4\textwidth}
         \centering
         \includegraphics[width=\textwidth]{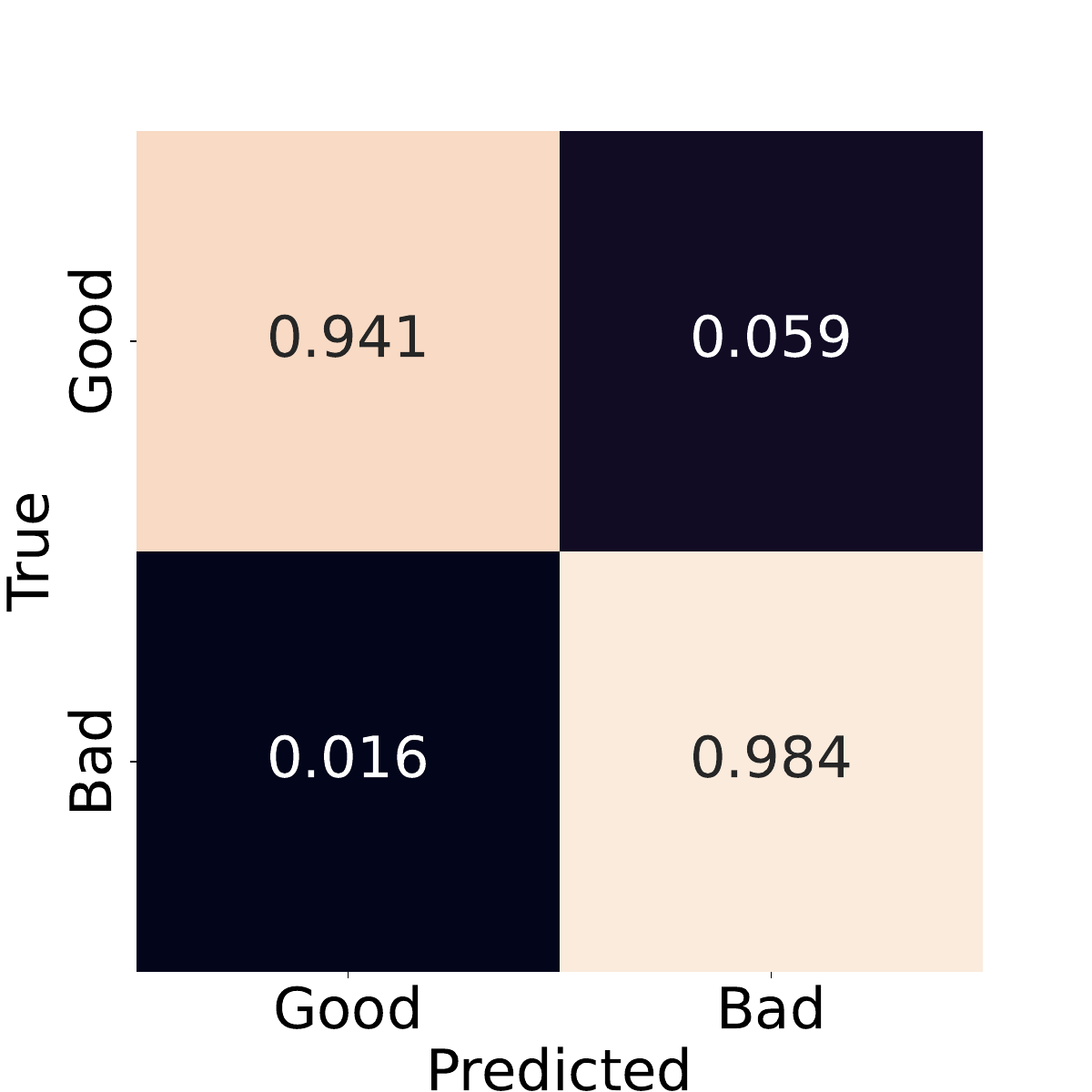}
         \caption{Image quality classifer CM}
         \label{fig: quality classifier CM}
     \end{subfigure}
     \begin{subfigure}[b]{0.4\textwidth}
         \centering
         \includegraphics[width=\textwidth]{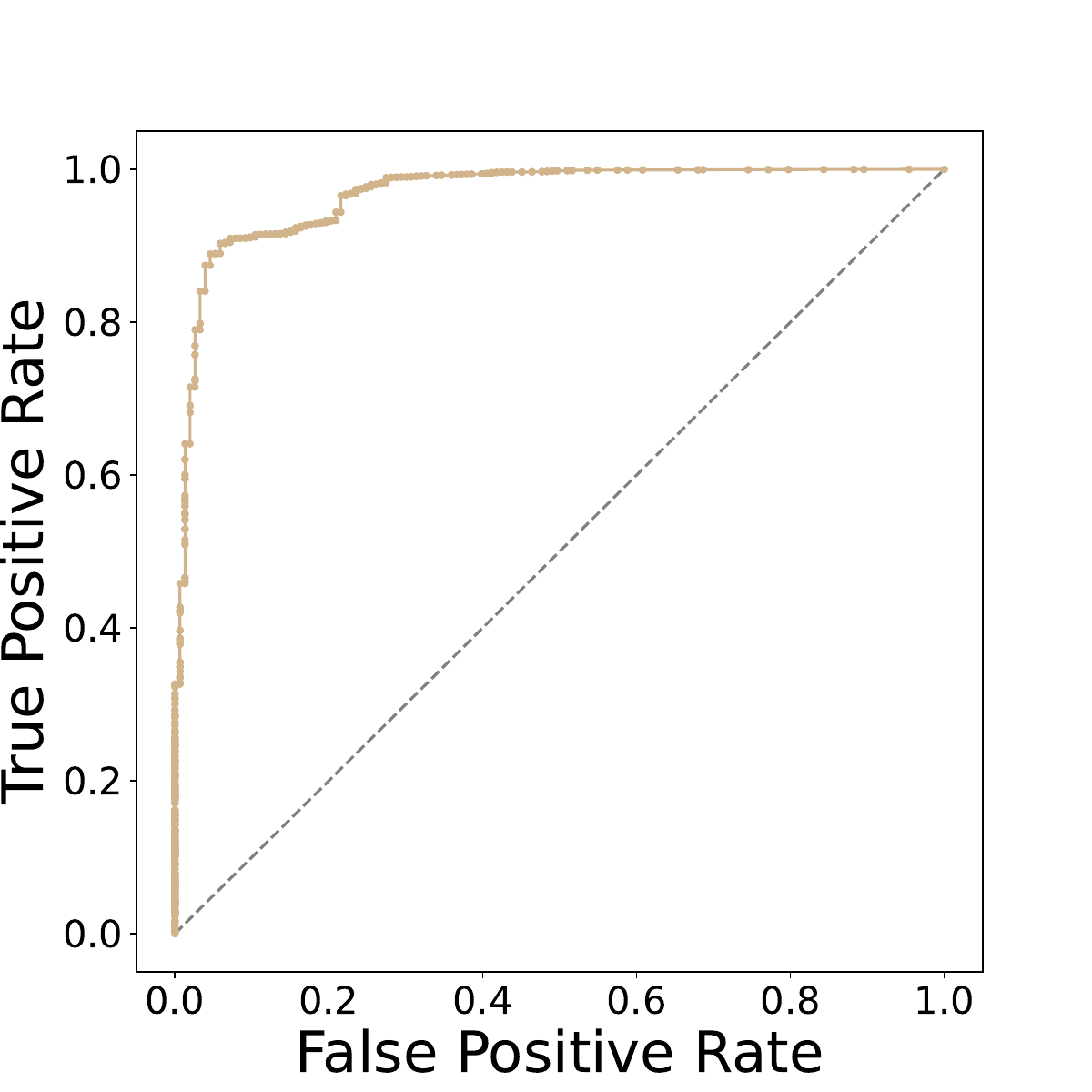}
         \caption{Image quality classifer ROC}
         \label{fig: quality classifer ROC}
     \end{subfigure}
        \caption{Results of our image quality classifier to reject images with large numbers of missing pixels. An image is assigned a binary classification as `Good' or `Bad'.}
        \label{fig:Quality classifier performance}
\end{figure*}

Firstly, we label those image with $\sigma> 1000$ as `bad', and manually select some examples of these bad images (in $g$ and $r$ bands). We label the others as `good'. Then we feed them into a simple two-layer CNN classifier for training and testing. The outputs give the probability of being a good image, shown in Figure \ref{fig: bad image samples}. Those good-quality images with a confidence greater than 0.5 are allowed for further processing, and those bad images are excluded.
Figure \ref{fig: quality classifier CM} and Figure \ref{fig: quality classifer ROC} show the confusion matrix and Receiver operating characteristic curve (ROC). The closer the curve is to the upper left corner, the more accurate the classifier is. 
The model rejects 98.4\% of bad images, and so we apply it as the first stage of data preprocessing process. In following experiments, about 12 peak images of ZTF objects are removed, taking 0.22\% of the whole image set.

\subsubsection{Z-scaling and normalization}

Astronomical pixel data can span a large dynamic range within a single image, which can cause problems for classifiers that need to learn faint features.
The IRAF Z-scale algorithm\footnote{\url{https://js9.si.edu/js9/plugins/help/scalecontrols.html}}, designed for displaying images as pixel intensity maps, is widely used to pick out features close to the background level of the image. The algorithm determines a minimum (z-min) and maximum (z-max) pixel value to display (pixels with values outside this range are displayed with the intensity at zero or saturated).

In our case, we apply the same z-scaling algorithm to replace any \textit{NaN} values or anomalously faint pixels with the z-min value. We do not apply a mask for z-max, to avoid treating real features (such as a bright transient) as saturated. Min-max normalization is then applied to the scaled data, limiting the values to [0,1].

\subsubsection{Data augmentation}

Data augmentation is a technique to create additional artificial samples within a training set. This is particularly helpful when dealing with classes containing few examples, such as our SLSNe and TDEs. Augmentation techniques for images include resizing, random flipping (horizontally or vertically), and random rotation (between 0 and 360 degrees, with any missing data at the edges filled with neighboring values). It is developed as a custom layer built after the input layer for convenience. 
While training, the images will be randomly modified through this layer for each epoch. 
Flips and rotations mean that the model is not encouraged to incorrectly learn specific location or orientation features. We do not apply resizing, in order to preserve the pixel scale of the data.

\subsection{Metadata preprocessing}

Metadata consists of the light curve features and the host galaxy magnitudes, colours and offsets. Details are shown in Table \ref{tab: metadata}. Currently, any missing metadata is replaced with zeros. 
Although a magnitude, time difference or offset of zero does have a physical meaning in this case, we find that adding zeros doesn't influence classifier performance in our experiments. Alternative methods will be considered in the next version of \textit{NEEDLE}.

Data standardization is applied for data scaling. Every feature is assumed to follow a Gaussian distribution among all samples, and individual values are scaled by its mean and standard deviation. In this way, the model can learn different feature distributions of the three classes, individually. 
Such scaling data are stored with the model.

\subsection{Data compression and indexing}

In order to feed a large amount of pre-processed data into the classifier for training on any computing platform, one convenient method is to store and fetch the data with HDF5 binary data format \citep{collette_python_hdf5_2014}. This allows users to transfer data among different facilities easily, and accelerate the training time for parameter optimization. 
In addition, a custom index has been added to each sample participating in training and testing, which can help users easily trace their ZTF IDs, thereby assisting case studies.
Here we store the image set, metadata set, labels, and sample index set in HDF5 format. Training/test set separation is conducted after loading the HDF5 data.

\section{Classifier Architecture and training}\label{sec: architecture}

In this section, we introduce the design of our \textit{NEEDLE} code and discuss the details of the model architecture.

We build our model within the \textit{Tensorflow Keras} framework.
We implement a custom Class called \textit{NEEDLE} that inherits from \textit{Keras.Model}. This Class includes the basic user-defined model functions (train, test, build, predict, loss function, etc), as well as model plotting and model visualization.

\subsection{Hybrid neural network}
To fully utilize the image and metadata, a hybrid model is required. Inspired by \citet{Carrasco_Davis_2021}, we build up a model that involves a block of convolutional layers for image inputs and a block of fully-connected layers for metadata inputs. 

\begin{figure*}
    \centering
    \includegraphics[width=\textwidth]{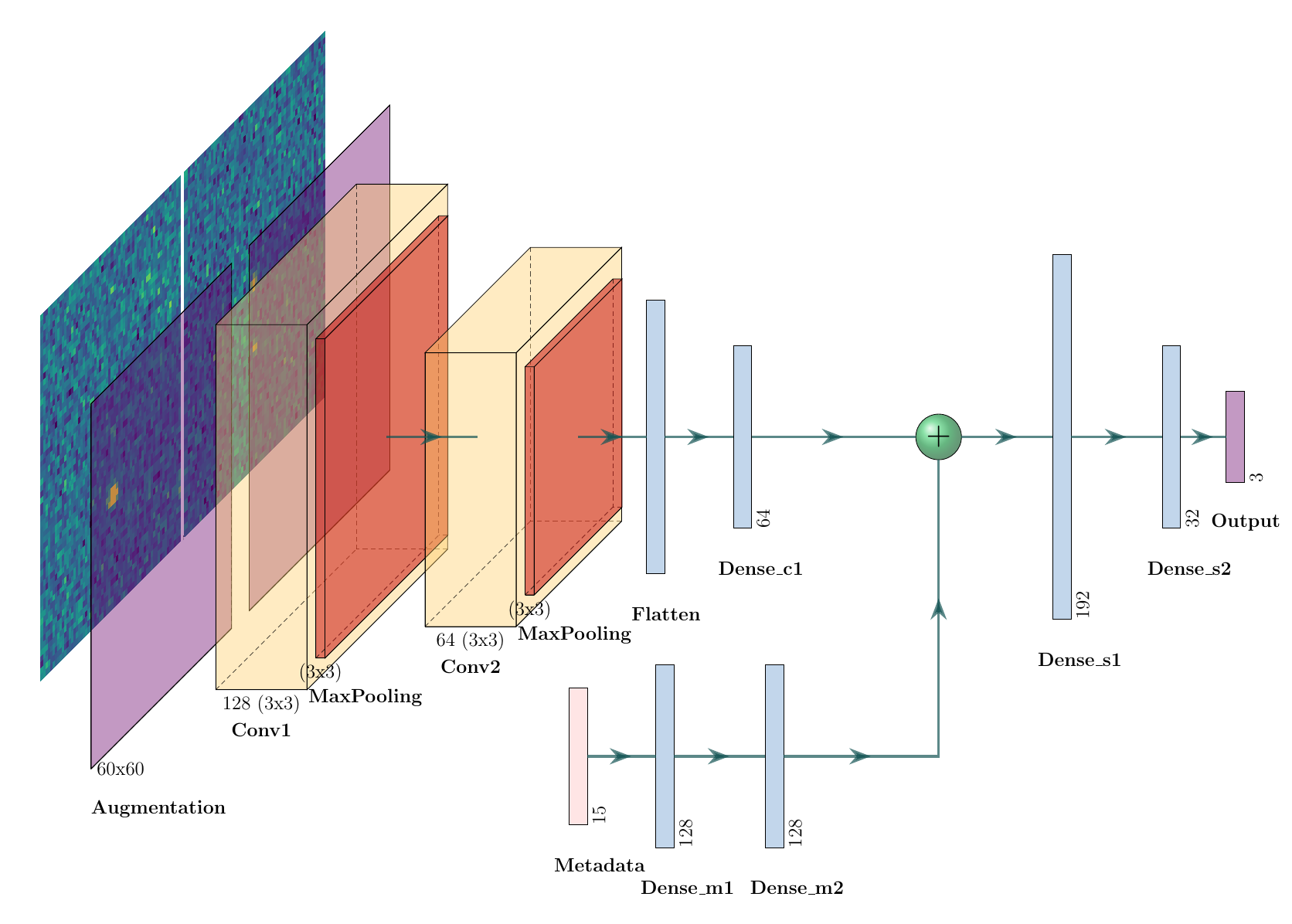}
    \caption{Model architecture for the full \textit{NEEDLE} classifier. The only difference between the \textit{NEEDLE-T} and \textit{NEEDLE-TH} variants is the length of the metadata input.}
    \label{fig:model architecture}
\end{figure*}

Figure \ref{fig:model architecture} shows the model architecture.
The image block consists of a data augmentation layer (random flipping and rotations) and two convolutional layers, each followed by a MaxPooling layer. The output of the last pooling layer is flattened into a 1D vector and fed into a fully connected dense layer with 64 neurons.
The metadata block consists of two fully connected dense layers with 128 neurons each.
The two types of outputs are then concatenated and fed into two dense layers (192, 32 neurons, respectively). Finally the outputs are fed into the output layer. 

Each layer uses a \textit{ReLU} acitivation function. The exception to this is in the final output layer, which uses a \textit{softmax} activation function to provide the probablities that an object belongs to each class.

\subsection{Training and test sets}\label{sec: train and test set}

Since the samples of SLSNe and TDEs are very small compared to the large number of normal SNe, training becomes difficult when more than $\sim20\%$ of them are put into the test set. Through experiments, we found that some objects are easily classified correctly with high probability, regardless of whether they are in the training set or the test set, however, some objects are difficult and return poor predictions.
Therefore, a fair approach that still allows us to train our model on a reasonable number of objects is to give a unique random seed, shuffle the dataset and randomly select test objects, repeating this process and training the model several times to average the results.
We choose to include 15 SLSNe, 15 TDEs and 15 SNe in the test set each time, and repeat this process 10 times to calculate the average model performance.

\subsection{Weighted loss function}

We start by importing the loss function \textit{SparseCategoricalCrossentropy} from \textit{Keras}. This is designed for multi-class tasks.
As our training data have extremely unbalanced labels and the majority are SNe, the model will naturally learn more SN features to quickly decrease the loss function, resulting in poor predictions for other classes. One solution is that we give more weight to rare labels and less weight to common labels. In this way, our model can extract features of different classes equally. Our weighted loss function is

\begin{equation}
    loss = -\frac{1}{N}\sum_{m=1}^{M}W_{m}\sum_{i=1}^{N_{m}}[y_{i}\log(\hat{y_{i}}) + (1-y_{i})\log(1-\hat{y_{i}})],
    W_{m} = \frac{1}{n_{m}}
\label{eq: loss}
\end{equation}

Here $N$ is the number of objects in one batch/epoch. 
$N_{m}$ and $n_{m}$ mean the number of objects from the $m$-th class in the batch/epoch and in the whole training set, respectively. $W_{m}$ is the class weight for the $m$-th class. $y_{i}$ and $\hat{y_{i}}$ refer to the true class and the model prediction for one input, respectively.

\subsection{Training optimization}

To set the learning rate, we employ the \textit{ExponentialDecay} method, which decreases the learning rate exponentially with growing steps while training. 
Equation \ref{eq: exponentialDecay} shows the algorithm. $lr_{0}$ is the initial learning rate. $\alpha$ means the user-defined decay rate. $N$ and $n$ mean the training step and user-defined decay steps, respectively. Through experimentation, we find acceptable performance using the following parameters: $l_{r_0} = 0.0002$, $\alpha = 0.95$ and $n = 100$.

\begin{equation}
l_{r_i} = l_{r_0} \times  \alpha ^{N / n}
\label{eq: exponentialDecay}
\end{equation}

An optimizer is the strategy for updating the weights and biases of neural networks, in order to help reduce the loss function to the desired minimum. For this model, we apply Adaptive Moment Estimation \citep[\textit{Adam};][]{Kingma_2014}, which is a special stochastic gradient descent algorithm that updates weights using two exponential decay rates. 

Overfitting is a non-negligible problem when training, which means that the model extracts noise\footnote{This is a particularly important problem for astronomical data as they are inherently noisy} rather than real features from the training data, resulting in high accuracy on the training set but low accuracy on the validation and test sets. To avoid this, \textit{tensorflow.keras.callbacks.EarlyStopping} is called to monitor the training process. When the loss for the validation set is not smaller than that at the previous 3 epochs, the training will stop. 

\subsection{Optimal network architecture}

We have tried and adjusted a variety of architectures and parameters. Given the size of the training set and the limited information in the images, a deep network is not suitable as it will likely lead to rapid overfitting and large fluctuations in the loss function. We therefore experiment with networks with only a few CNN layers.

\textit{KerasTuner} \citep{omalley_2019} is an easy-to-use, scalable hyperparameter optimization framework that allows users to set ranges of neurons, activation function, and learning rates. It will automatically run every combination of configurations and search for optimal solutions. 
We apply this method to adjust the architecture and hyperparameters in \textit{NEEDLE}.

The results show that a model with two Convolutional layers (each 128 3*3 kernels) for image inputs, and two fully-connected layers (each 64, 128 neurons) for metadata inputs, is able to perform the best predictions. The detailed architecture is shown in Figure \ref{fig:model architecture}. The learning rate, batch size and number of epochs are $3e^{-5}$, 128, and 300, respectively.

\begin{figure*}
        \begin{subfigure}[b]{0.32\textwidth}
            \centering
            \includegraphics[width=\textwidth]{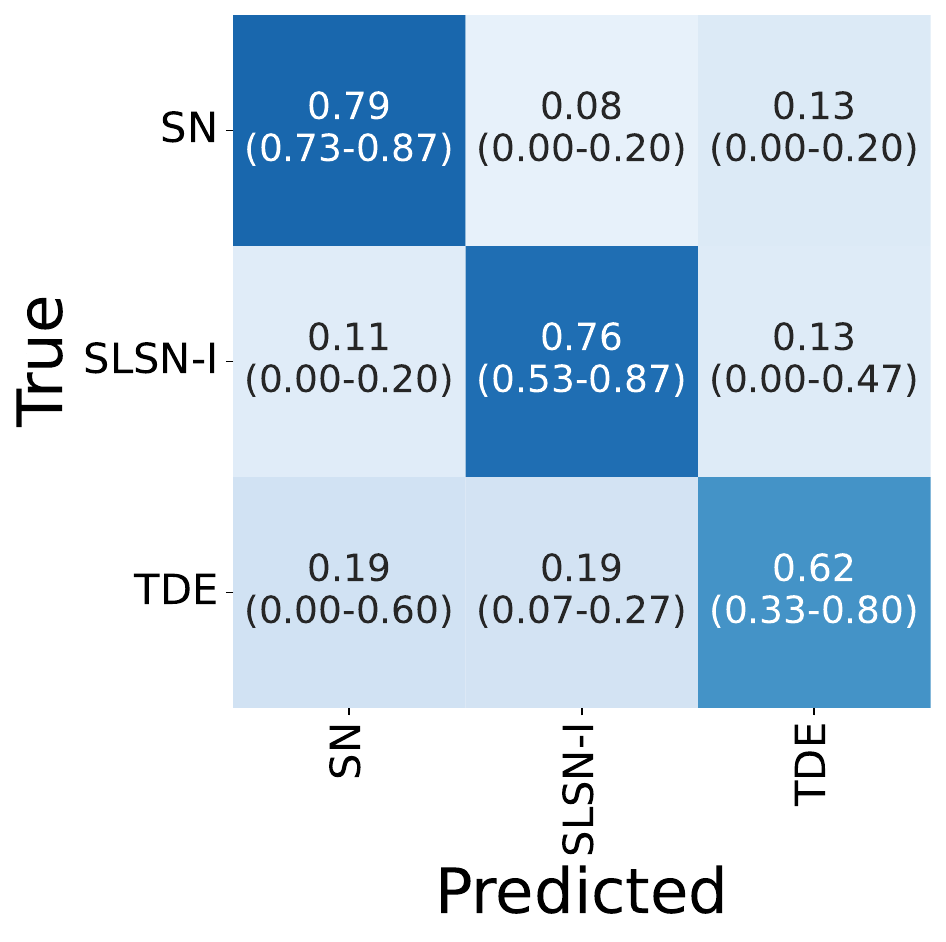}
            \caption{ \textit{NEEDLE-T} Classifier, trained using images and transient light curve features.}   
            \label{fig: cm v1}
        \end{subfigure}
        \hfill
        \begin{subfigure}[b]{0.32\textwidth}  
            \centering 
            \includegraphics[width=\textwidth]{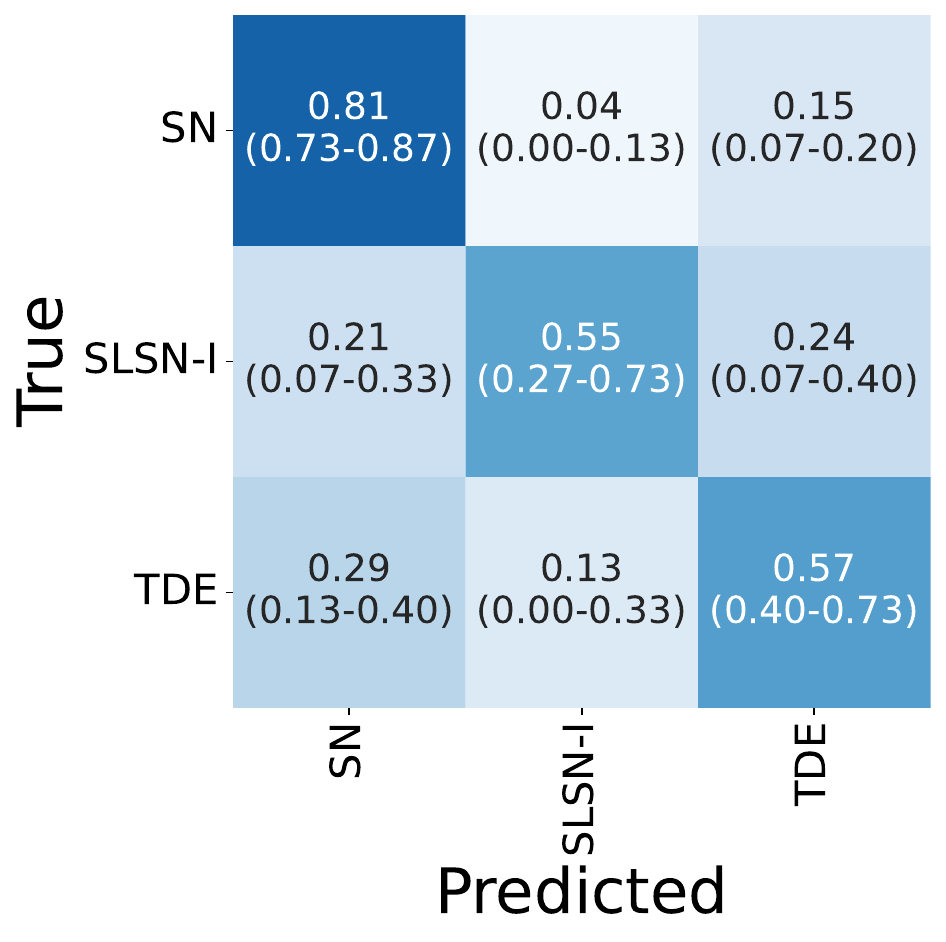}
            \caption{ \textit{NEEDLE-T} Classifier when trained using only those objects with cataloged hosts.}   
            \label{fig: cm v1 with hosts}
        \end{subfigure}
        \hfill
        \begin{subfigure}[b]{0.32\textwidth}   
            \centering 
            \includegraphics[width=\textwidth]{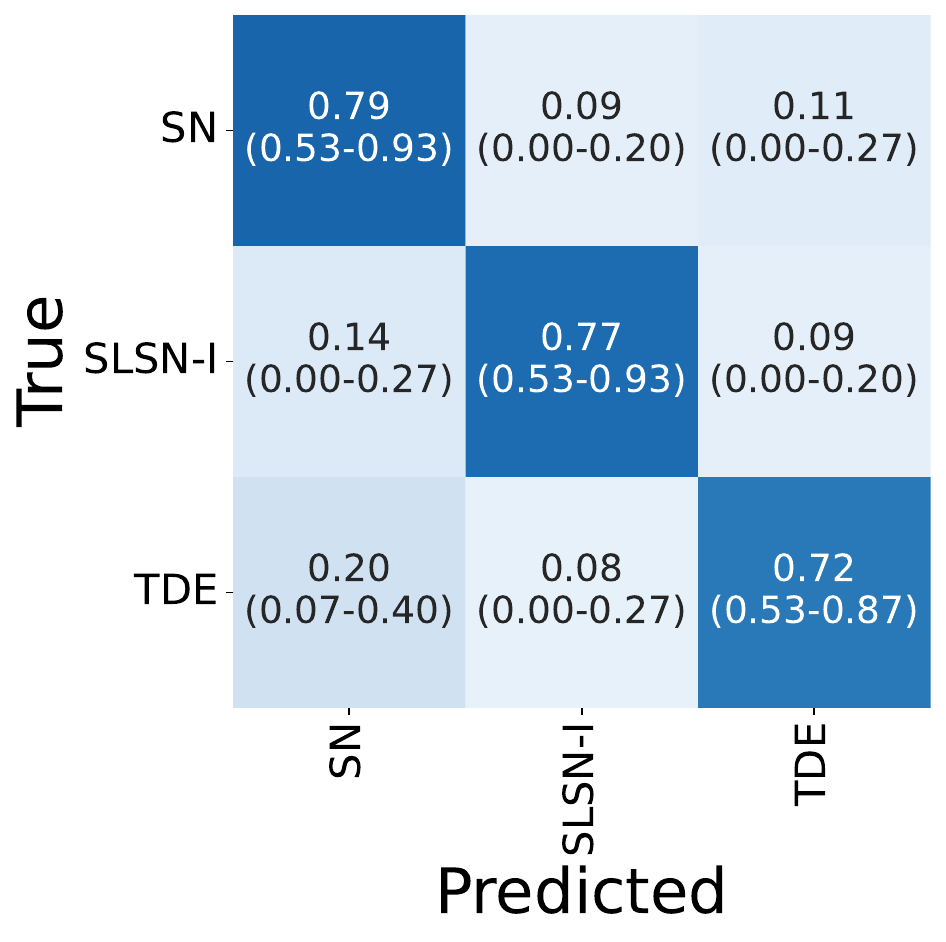}
            \caption{ \textit{NEEDLE-TH} Classifier, trained using images and all transient and host metadata.}   
            \label{fig: cm v2}
        \end{subfigure}
     
        \caption{Confusion matrices of \textit{NEEDLE-T} and \textit{NEEDLE-TH} (without and with host photometry) classifiers on an unseen test set. For each classifier, the random seed for initializing parameters is unique, and the test/training sets are randomly shuffled before training. The values reported in each confusion matrix are the averages for 10 realisations of this process, and the ranges across all 10 models are shown in brackets.}
    
        \label{fig:cm_plots}
    \end{figure*}

\section{Experiments and results}\label{sec: results}
In this section, we investigate model performance on the ZTF BTS sample. In particular, we aim to determine which metadata are important to include in our training and test sets, the expected purity and completeness, and how confidently we can predict the type of an object at early phases. We also present the \textit{NEEDLE} pipeline that we are implementing on \textit{Lasair}.

\begin{figure*}
    \centering
    \includegraphics[width=\textwidth]{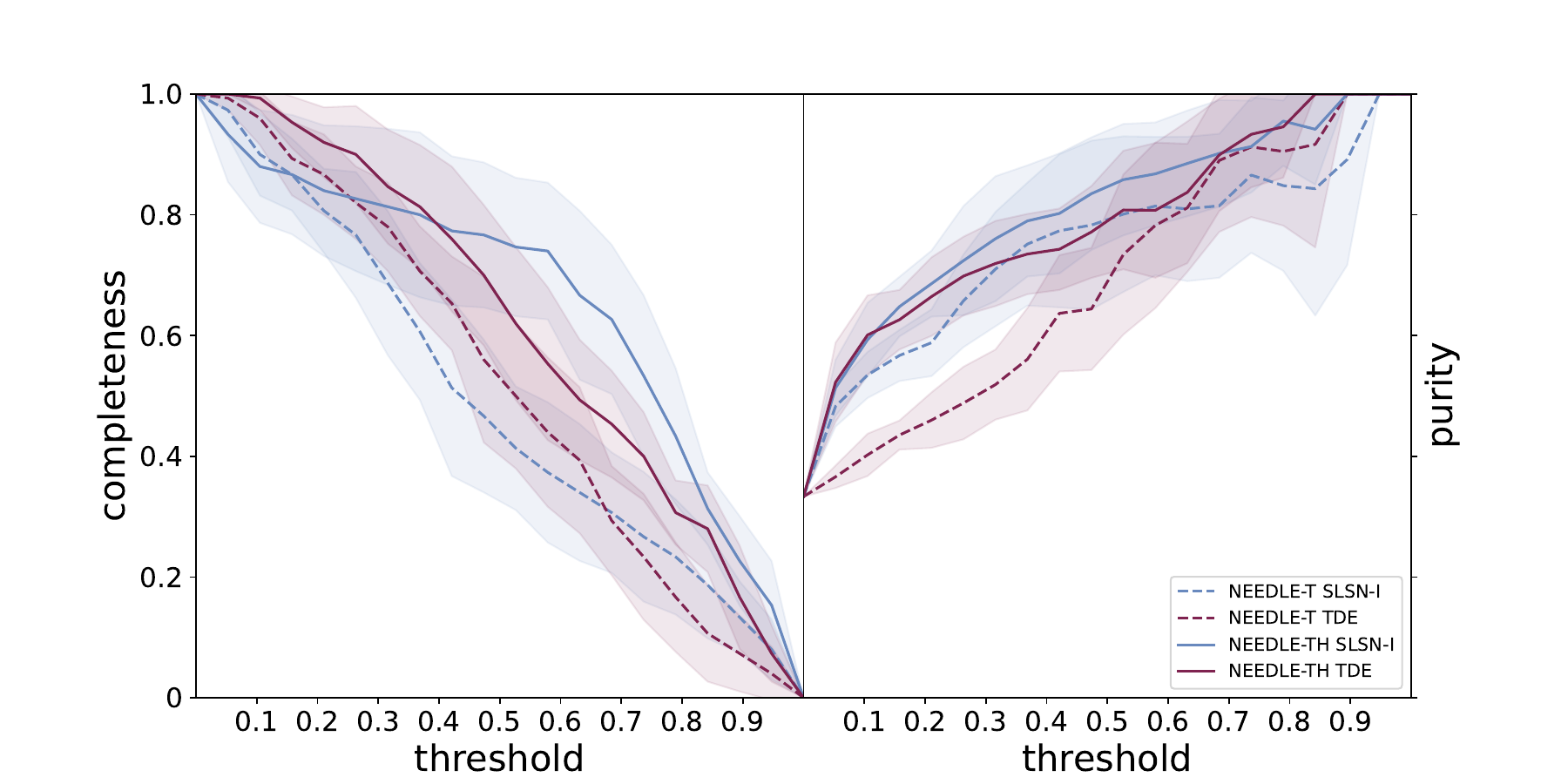}
    \caption{Completeness and Purity for \textit{NEEDLE-T} and \textit{NEEDLE-TH} classifiers. The results are averaged among 10 models with 10 random test sets. }
    \label{fig:completeness_purity_plots}
\end{figure*}

\begin{figure*}
        \begin{subfigure}[b]{0.32\textwidth}
            \centering
            \includegraphics[width=\textwidth]{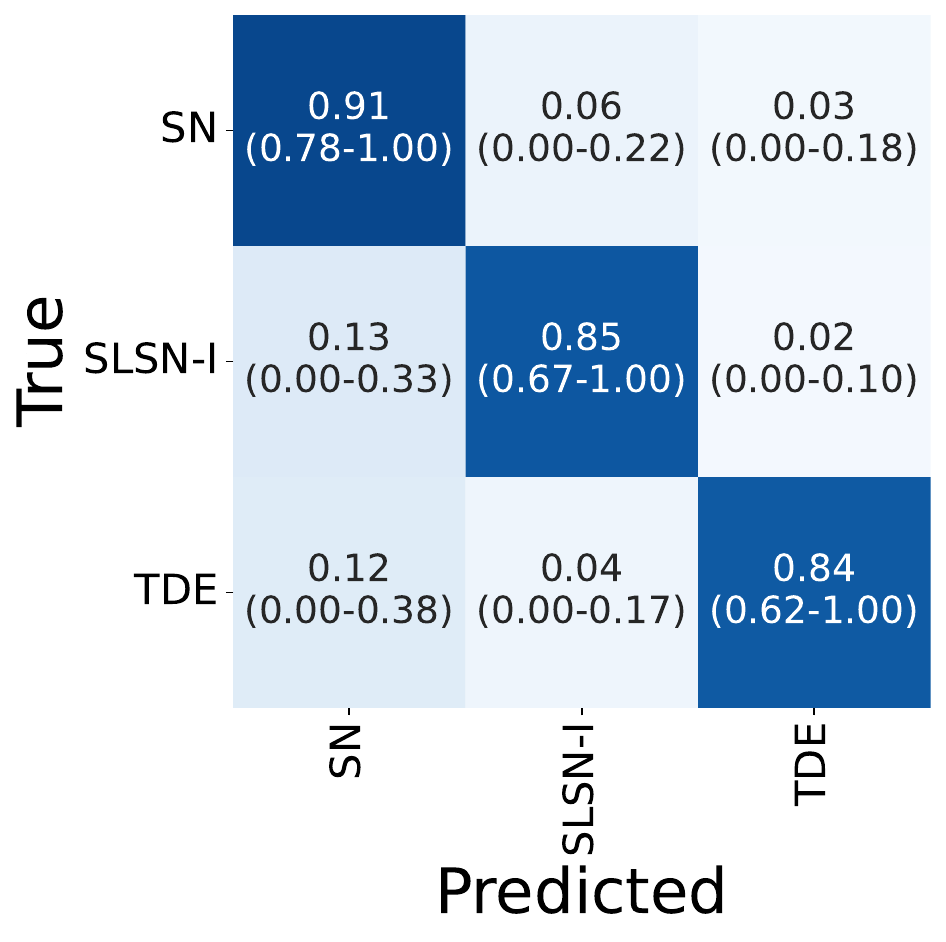}
            \caption{NEEDLE-TH completeness on balanced test set.}   
            \label{fig: NEEDLE-TH in test set}
        \end{subfigure}
        \hfill
        \begin{subfigure}[b]{0.32\textwidth}  
            \centering 
            \includegraphics[width=\textwidth]{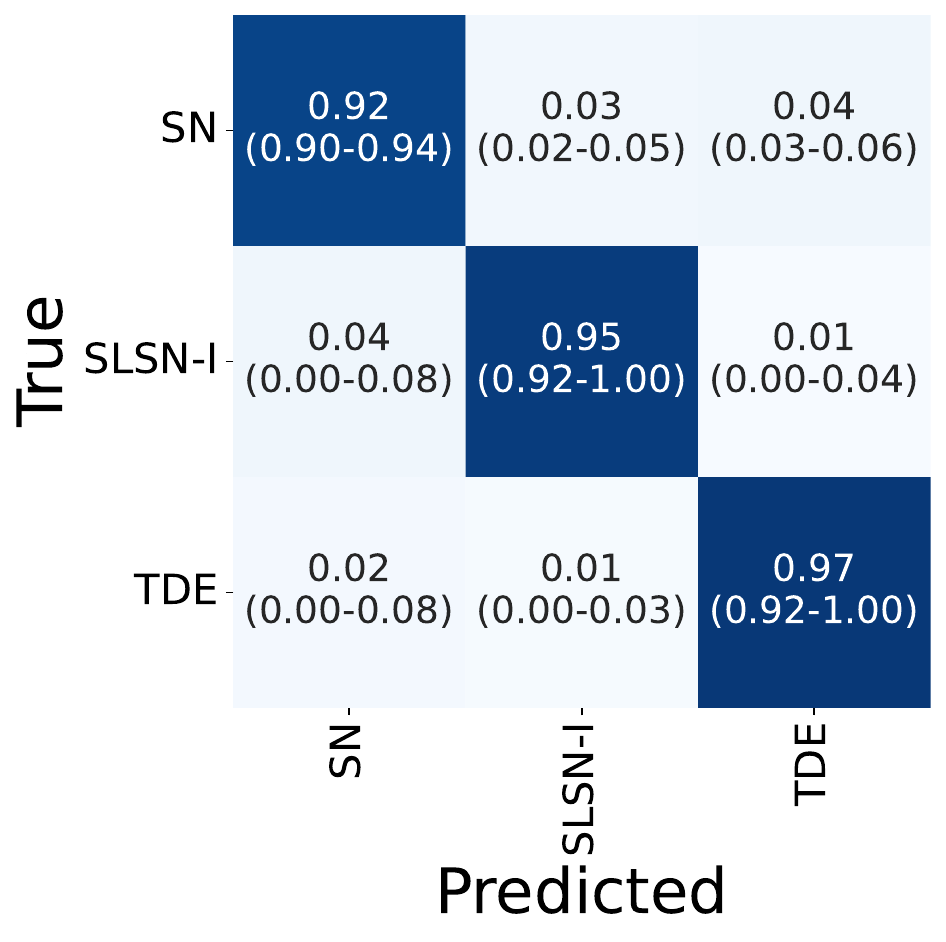}
            \caption{NEEDLE-TH completeness on full data set.}   
            \label{fig: NEEDLE-TH in full set}
        \end{subfigure}
        \hfill
        \begin{subfigure}[b]{0.32\textwidth}   
            \centering 
            \includegraphics[width=\textwidth]{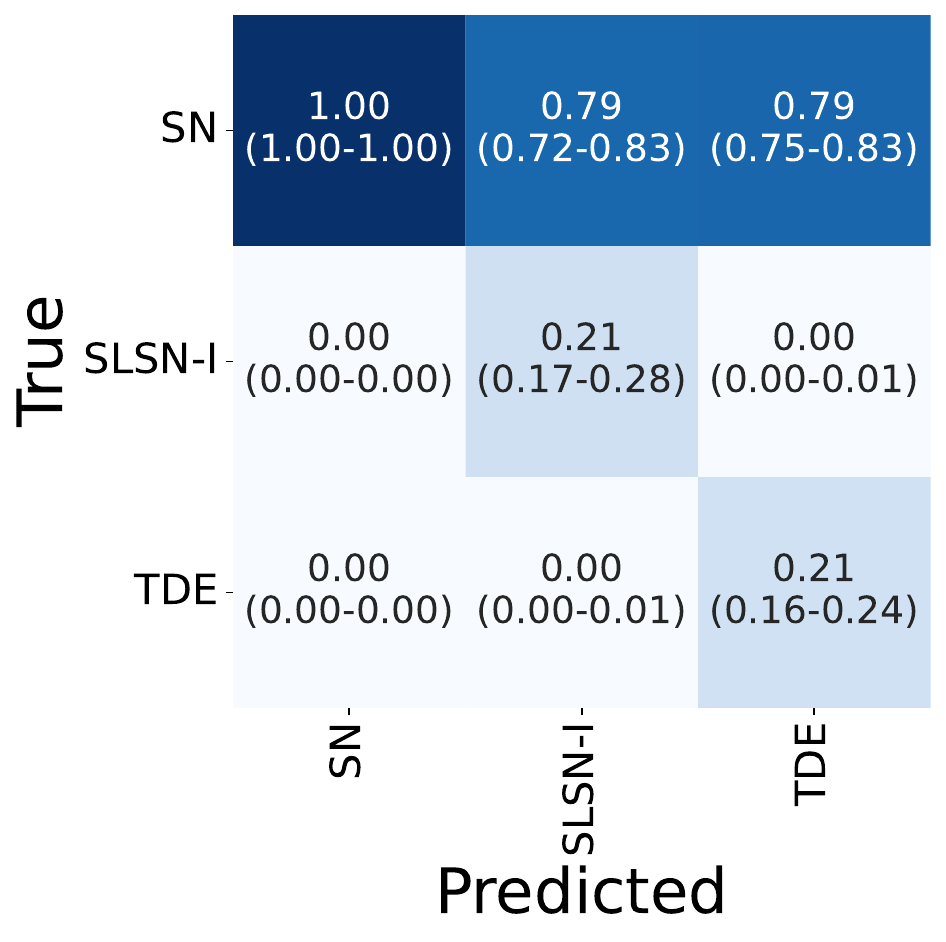}
            \caption{NEEDLE-TH purity on full data set.}   
            \label{fig: NEEDLE-TH purity in full set}
        \end{subfigure}
     
        \caption{Confusion matrix (CM) of \textit{NEEDLE-TH} restricted to objects with classification confidence $p({\rm class})>0.75$, for the test set and the full data set. The values reported in each confusion matrix are the averages after training 10 times with randomly shuffled test and training sets, and the ranges across all 10 models are shown in brackets. The first two CMs depict the completeness in the test set (Figure \ref{fig: NEEDLE-TH in test set}) and the full set (Figure \ref{fig: NEEDLE-TH in full set}). The third CM (Figure \ref{fig: NEEDLE-TH purity in full set}) shows the purity of each class in the full set, which is representative of the expected balance in real-time alerts.  }
    
        \label{fig: cm in diff sets}
    \end{figure*}

\subsection{Classifier performance with \& without host metadata}
Initially, we train only with information available from the real-time transient alerts: the \textit{science} image (here assumed close to the time of maximum light), the \textit{reference} image, and the transient metadata such as magnitude and time since discovery. For convenience we call this version \textit{NEEDLE-T} (for transient).
We then retrain the model, this time including the cataloged host galaxy properties obtained from \textit{Sherlock} and Pan-STARRS, and we label this version \textit{NEEDLE-TH} (transient+host). 

Figure \ref{fig:cm_plots} shows the confusion matrix with the completeness of the three models on the test set. The prediction is decided by the maximal probability among three classes. The values given in the confusion matrix are the averages of 10 model realisations with randomly shuffled test sets (containing 15 objects per class each time, with remaining objects in the training set).
The initial \textit{NEEDLE-T} classifier (Figure \ref{fig: cm v1}) can recognize 79\% of normal SNe and 76\% of SLSNe in the test set on average. It is worth recalling that more than half of SLSNe in our sample do not have cataloged hosts, therefore only 41/87 SLSNe can be included in the full \textit{NEEDLE-TH} model. If we train \textit{NEEDLE-T} only on these objects with detected hosts (enabling a fair comparison later with \textit{NEEDLE-TH}), the averaged true positives of SLSNe decrease slightly to 55\%, and the large range shows that the predictions are less stable with the smaller sample. This is shown in Figure \ref{fig: cm v1 with hosts}. Adding the host metadata in \textit{NEEDLE-TH} (Figure \ref{fig: cm v2}) improves the performance for SLSNe to 77\% on average, despite the smaller sample size, showing the importance of including host galaxy information. 
Even in the worst-performing model, at least 53\% SLSNe are correctly identified with the help of host magnitudes and colour information, and the highest completeness reaches 93\%.

This effect is even more pronounced for TDEs.
For \textit{NEEDLE-TH}, the average true positive rate for TDEs grows from 57\% to 72\% with the addition of host information. While galaxy colours do differ for TDEs compared to the other transient types, more likely this improvement reflects the fact that all TDEs in the sample have a small offset because they occur in the nuclei of their hosts.

\subsection{Completeness and Purity}

Figure \ref{fig:completeness_purity_plots} shows the completeness and purity trends on the unseen test set with increasing probability thresholds for classification, in both \textit{NEEDLE-TH} (corresponding to Figure \ref{fig: cm v2}) and \textit{NEEDLE-T} (for only those objects having cataloged hosts, corresponding to Figure \ref{fig: cm v1 with hosts}).  Here a class is only assigned if $p({\rm class})>x$ for the most probable class. In each case we show the average and standard deviation of 10 trained models.

We can see that on average, for SLSNe we attain a completeness 76\% (52\%) and a purity 85\% (92\%) for a threshold $p({\rm SLSN})>=0.5 (0.75)$. For TDEs, we obtain completeness 66\% (38\%) and purity 80\% (93\%).
The results are fairly competitive with other popular classifiers (shown in Table \ref{tab: comparison with classifiers}, especially considering that we only use single images and limited light curve information.

We note that the purity achieved with our balanced test set will likely not reflect the purity obtainable in a real survey, due to the large imbalance in rates between SLSNe/TDEs and normal SNe. Therefore, when selecting objects in real time, one may wish to choose a high probability threshold to minimise the absolute numbers of normal SNe mis-classified as SLSNe or TDEs.

Figure \ref{fig: cm in diff sets} shows confusion matrices for transients classified with probability $p({\rm class})>=0.75$. We show the completeness for \textit{NEEDLE-TH} on the balanced, unseen test sets (Figure \ref{fig: NEEDLE-TH in test set}) and completeness and purity matrices for the full data set (Figures \ref{fig: NEEDLE-TH in full set}, \ref{fig: NEEDLE-TH purity in full set}). With $p({\rm class})>=0.75$, \textit{NEEDLE-TH} can correctly classify 95\% TDEs and 97\% SLSNe-I in the full data set. However, for even just a few \% SN contamination, this results in a real-world purity of around 20\% for the rare classes, showing the importance of choosing a probability threshold carefully. \textit{NEEDLE} is designed to select young SLSN and TDE candidates for spectroscopic follow-up, rather than to produce large photometric samples. Therefore, a purity of a few $\times10\%$ is an acceptable price for the high completeness.

We also investigate the importance of including host galaxy metadata. The diagrams show that SLSNe and TDEs essentially always gain higher completeness and purity when host metadata is included.

\subsection{Classification from early detections}\label{probs before peak}
\begin{figure*}
\centering

\begin{subfigure}{\textwidth}
    \centering
    \includegraphics[width=\linewidth]{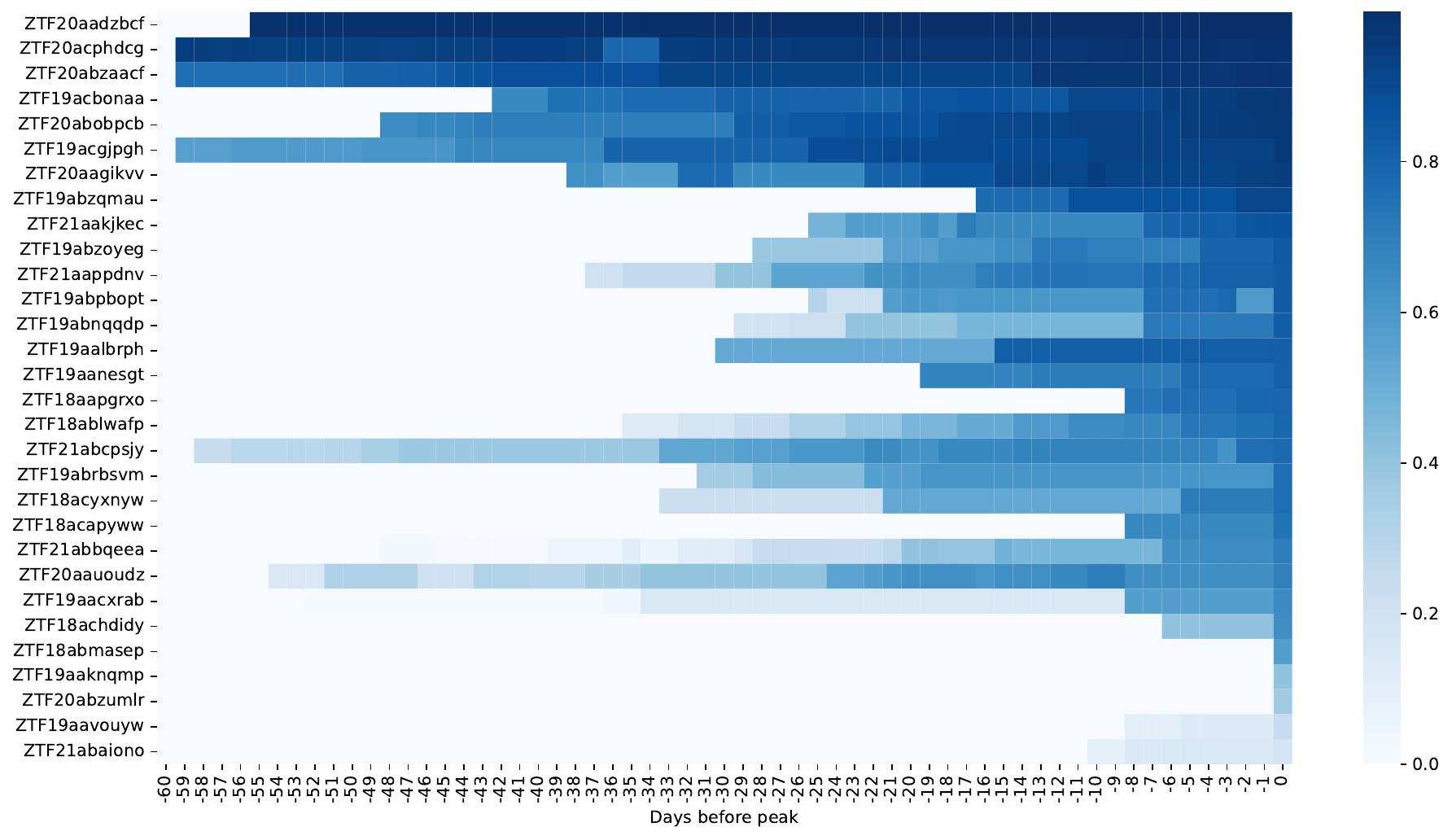}  
    \caption{SLSNe}
    \label{fig: SLSN trends}
\end{subfigure}
\begin{subfigure}{\textwidth}
    \centering
    \includegraphics[width=\linewidth]{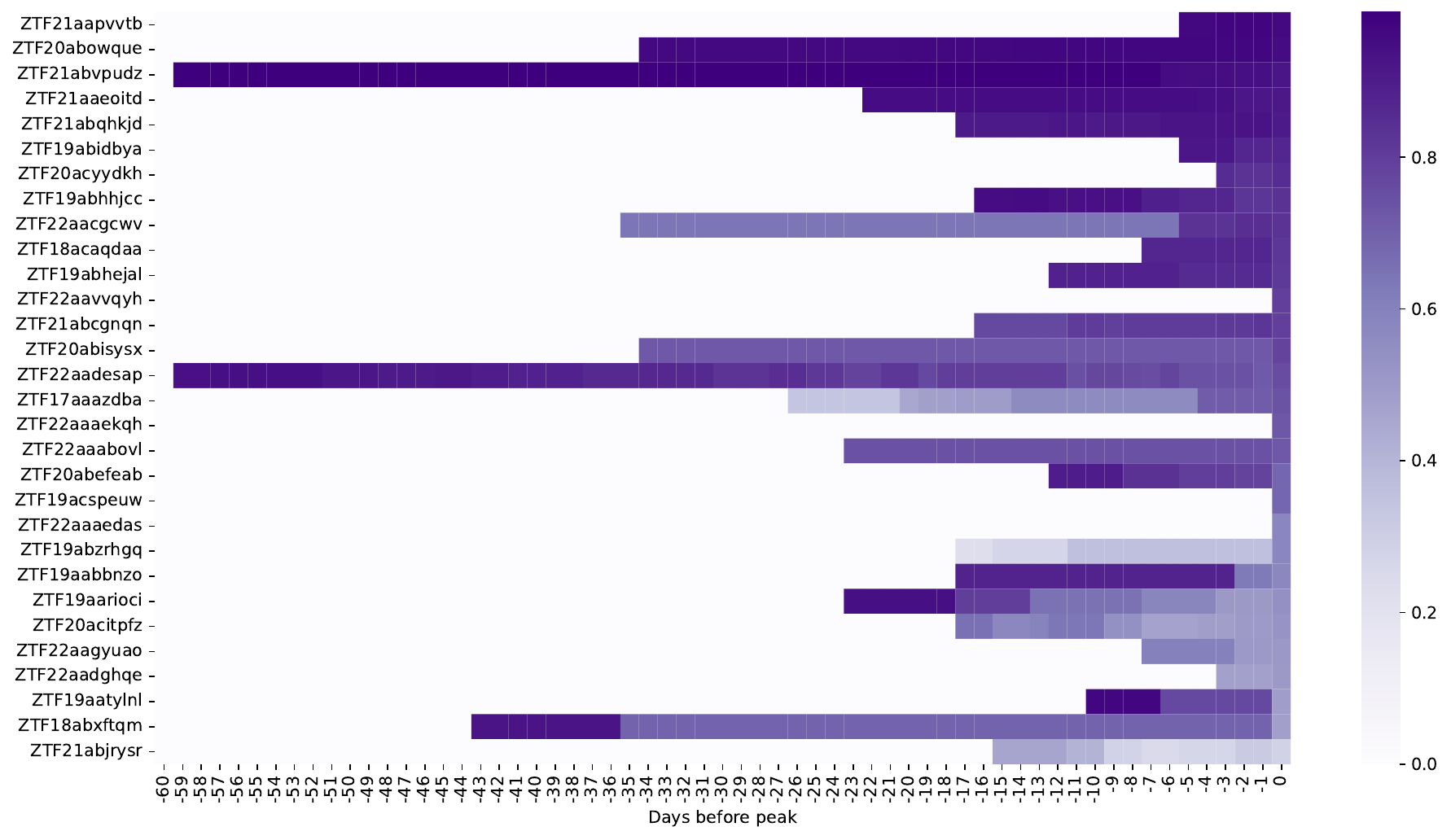}  
    \caption{TDEs}
    \label{fig: TDE trends}
\end{subfigure}
\caption{Probability heatmaps for SLSNe-I (\ref{fig: SLSN trends}) and TDEs (\ref{fig: TDE trends}). In each class, 30 objects are randomly selected. The x-axis is the date starting from 60 days before the peak of the event to the peak date. The y-axis is the ZTF objects' names. The colour bar corresponds to the probability range. ZTF objects are sorted in descending order of their peak probability.}
\label{fig: label probs trends}
\end{figure*}

As \textit{NEEDLE} is designed to provide a probability for each label after only a few early detections, we also test the average performance of \textit{NEEDLE-TH} over time since explosion by attempting to classify a time series of pre-peak detections of 30 randomly selected objects in each class. We show the predicted $p({\rm SLSN})$ against time before peak for 30 SLSNe, and $p({\rm TDE})$ for 30 TDEs, in Figure \ref{fig: label probs trends}. 

For most SLSNe (Fig \ref{fig: SLSN trends}) and TDEs (Fig \ref{fig: TDE trends}), the probability assigned to the correct class grows as the events approach the peak. This is likely due to the longer baseline over which the light curve features can be evaluated, indicating that properties such as light curve rise time and slope and host galaxy contrast are important features in \textit{NEEDLE}. This is particularly apparent in the case of SLSNe, where magnitude contrast with the host galaxy (which is maximised at light curve peak) is also an important feature.

\subsection{Real-time annotation on Lasair}\label{sec: classification on lasair}

We aim to provide \textit{NEEDLE} classifications in close to real time via the LSST:UK alert broker, \textit{Lasair}. Our classifier will digest incoming transients from a pre-filtered \textit{Kafka} stream produced by a simple \textit{Lasair} query, using data from ZTF (or LSST in the future), and provide the probabilities of different classes for each object. To return our classifications to the broker, we make use of the \textit{Lasair} annotator\footnote{\url{https://lasair.readthedocs.io/en/main/concepts/annotations.html}} feature, which allows verified users to add information to the transients database in a format that is query-able by another user.

\begin{figure*}
    \centering
    \includegraphics[width=0.8\textwidth]{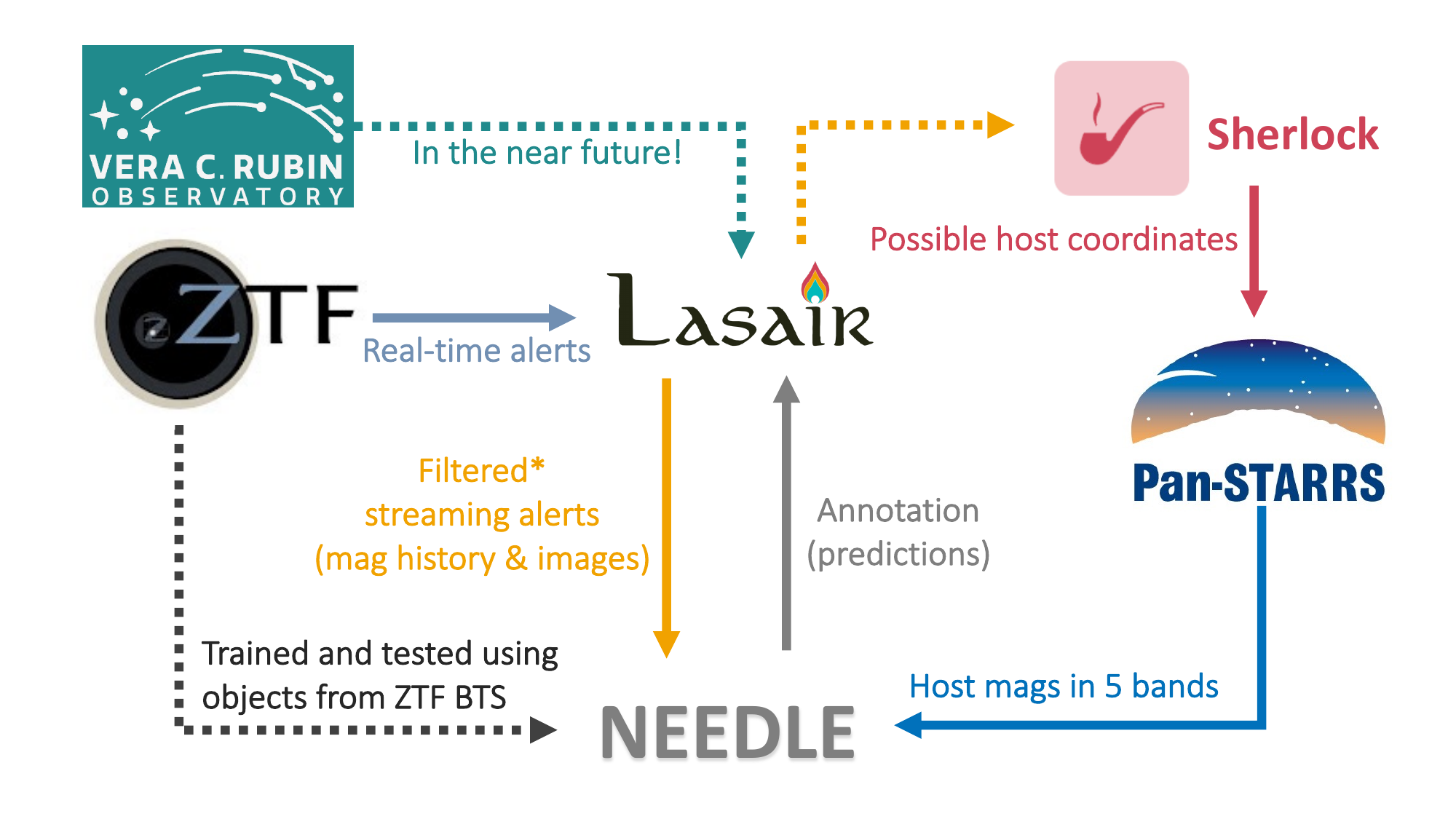}
    \caption{NEEDLE pipeline design for the alert broker \textit{Lasair}. NEEDLE receives ZTF (and ultimately LSST) alerts from \textit{Lasair} via a customized SQL filter to remove old or bogus objects. The \textit{Science} and \textit{Reference} images are contained in the ZTF alerts, or requested from the Rubin Science Platform. If they pass the quality checker, the host metadata will be fetched from \textit{Sherlock} and Pan-STARRS. Finally, \textit{NEEDLE} will return the probabilities for the three classes to the \textit{Lasair} annotation database, allowing them to be used in subsequent alert filters by any user.
    }
    \label{fig:needle_pipeline}
\end{figure*}

Figure \ref{fig:needle_pipeline} shows the process in detail.
\textit{NEEDLE} is trained and tested using the ZTF alerts coming from \textit{Lasair}.
New alerts will be filtered by a customized SQL query to provide only young, reliable, extragalactic, non-repeating transients. Specifically, we retain events: 
\begin{itemize}
    \item discovered within the last 60 days
    \item with more than 3 confident detections (to reduce the chance of bad subtractions)
    \item predicted to be a Supernova or Nuclear Transient by \textit{Sherlock} (i.e. not a known AGN or Galactic variable).
\end{itemize}
Then, \textit{NEEDLE} selects the brightest available detection as the input image, if it passes the quality image checker. \textit{NEEDLE} then collects the host coordinates and photometry from \textit{Sherlock} and Pan-STARRS, computes the predicted probabilities from the trained network, and sends them back to \textit{Lasair} as annotations.

We have tested this process end-to-end with a preliminary version of \textit{NEEDLE}. Our goal is to run the fully trained \textit{NEEDLE} model automatically on all ZTF alerts passing our SQL filter, and release the results as a public stream on \textit{Lasair}, beginning in early 2024.

\section{Discussion} \label{sec:discussion}

\subsection{Individual discrepancy among rare transients}
As mentioned in Section \ref{sec: train and test set}, the difficulty of classifying each individual object in our data set varies. 
One reason for this may be issues with the host galaxy metadata. In the Pan-STARRS survey, very nearby resolved galaxies may be broken into multiple sources by the survey photometry pipeline, resulting in underestimated host magnitudes. Failed host association may also cause issues, leading to the wrong photometry being retrieved. This is a particular problem for SLSNe, where many of the true hosts are not detected. 

We also identify several real features of our objects that influence the ease of classification.
 For SLSNe, we found those that are easily classified often have relatively high $m_{\rm discovery}$ and $\Delta T_{\rm discovery}$, low $\Delta m_{\rm discovery}$, in a slightly bluer and faint (low ${g-r}$ and ${r-i}$) host galaxy, i.e. they are bright with a slow rise and a star-forming host, consistent with classic SLSNe in the literature. SLSNe in slightly more massive galaxies, or with short rise times, are more difficult to separate from normal SNe.

 For TDEs, objects are most easily classified if they have a bright $m_{\rm discovery}$ and a shorter $\Delta T_{\rm discovery}$ than typical SLSNe. This could also occur because these events occur in the nuclei of galaxies, and so tend to be found closer to peak unless the flux contrast with the host is large.  In future work we will investigate in more detail how to optimise the training process to account for these variations.

\subsection{Comparisons to previous classifiers}
In recent years, several transient classifiers have been designed that can recognize TDEs and in particular SLSNe. Some of them gain excellent accuracy for SLSNe by making use of their uniquely slow light curves. For the same reason, many of these classifiers show better performance when more light curve data are available at later phases.
Table \ref{tab: comparison with classifiers} shows comparisons of these classifiers with our \textit{NEEDLE} Classifier. 

The advantage of \textit{NEEDLE} is that we do not require multiple detections or host redshifts as input because at LSST depth, few galaxies will have spectroscopic redshifts.  We only use single-stamp images, alert photometry and cataloged host magnitudes (when available), enabling an informed real-time prediction from as little as one detection. Furthermore, all data used in training is from real survey detections rather than simulations. 
It is likely that we could gain an even better performance by making use of more detailed light curve information, and this is the aim for future development. However, the goal of \textit{NEEDLE} is not to produce pure samples of photometrically classified events, but to provide probabilities of potential SLSNe and TDE at an early stage to guide spectroscopic follow-up. From this perspective, completeness may be more important than purity.

\begin{table*}
\begin{adjustbox}{angle=90}
\begin{tabular}{p{40pt}|p{40pt}|p{30pt}|p{100pt}|p{150pt}|p{150pt}}

\hline
Code & Paper & Model & Data sources& Inputs& Performance for SLSNe \& TDEs\\ \hline

tdescore & \citet{Stein_2023} & XGBoost & ZTF alerts, Pan-STARRS hosts & 10 features of full light curves, 5 features from the context. & In balanced test sets: TDE completeness of 77.0\%, purity of 80.3\% .\\ \hline

FLEET-SLSN    &  \citet{Gomez_2020}  &   Random forests   & \textbf{Supernovae: }Open Supernova Catalog (OSC, \citet{Guillochon_2017}), ZTF; 

\textbf{Host: }SDSS, PS1/3$\pi$ & \textbf{Light curve features:} width of light curves, the phase offsets, the peak magnitudes from $g$ and $r$ bands; 

\textbf{Host features:} apparent magnitudes, half-light radius in $r$ band, offset (same as \textit{NEEDLE} here), offset normalized by galaxy radius, the apparent magnitude difference between transient at peak and host. & In unbalanced/observed sets: SLSN purity of about 85\% and completeness of 20\%  \\ \hline
FLEET-TDE    &\citet{Gomez_2023}  &   Random forests   &  \textbf{Transients: }spectroscopically classified transients from TNS, light curves from ZTF; 

\textbf{Host: }SDSS, PS1/3$\pi$  &  Similar to FLEET-SLSNe.
     & In unbalanced/observed sets: 20 days after discovery: about 40\% completeness and about 30\% purity; 

40 days after discovery: about 40\% completeness and about 50\% purity.         \\ \hline
ALeRCE-light curve classifier&  \citep{Sanchez-Saez_2021} &  Balanced random forests &  ZTF light curves, with labels from a variety of catalogues. &  \textbf{Detection features:} 56 features per band, and 12 features computed using $g$ and $r$ bands, yielding a total of 124 detection features;

\textbf{Non-detection features:} 9 features per band defined using all the non-detections associated with a given source. &   In balanced test sets: 100\% accuracy with 26\% deviation, high accuracy though only 24 SLSNe samples.   \\ \hline

SCONE & \citet{Qu_2022}     &     CNN       & A set of LSST deep drilling field simulations.          & 2D Gaussian process generating flux heatmaps as function of time and filter (wavelength).   & In balanced test sets: Without redshift, SLSN accuracy is 0.69, 0.76 and 0.92 at 0, 5 and 50 days after discovery; with redshift, the values are 0.91, 0.93 and 0.97.\\ \hline

SN classifier & \citet{Burhanudin_2023} &  CNN \& transfer learning  &  Light curves from Open Supernova Catalog; the Photometric LSST Astronomical Time Series Classification Challenge (PLAsTiCC). & Referred from  \citet{Qu_2022}&  In unbalanced test sets: for PLAsTiCC-simulated SLSNe, the accuracy with and without redshift are 0.61 and 0.65, respectively.        \\ \hline

RAPID  & \citet{Muthukrishna_2019}  &   RNN  & Simulated data from PLAsTiCC transient models. & A matrix with each row composed of the imputed light-curve fluxes for each band, repeated values of the host galaxy redshift, and the MW dust reddening.  &  In (nearly) balanced test sets: for SLSNe, the accuracy is 0.83 (2 days after trigger) to 0.85 (40 days); for TDEs, the numbers are 0.59 and 0.86.  \\ \hline

SuperRAENN &  \citet{Villar_2020}   &  Recurrent autoencoder (RAE) \& Random Forests    &    Light curves from Pan-STARRS1 Medium Deep Survey (PS1 MDS) with known redshifts.  &   Gaussian-processed light curve for RAE inputs, then use RAE latent features as inputs for random forest.      &   In unbalanced test sets: for SLSNe, the completeness and purity are 0.76 and 0.81, with threshold larger than 0.7, their values increase to 0.83 and 0.91, respectively. Host redshift is considered. \\ \hline

Superphot & \citet{Hosseinzadeh_2020} &      Random forests     &  Light curves from Pan-STARRS1 Medium Deep Survey (PS1 MDS)                  &  6 Principle Component Analysis coefficients on modelled light curves with known redshifts.      &  In unbalanced sets: for SLSNe, the completeness and purity are 0.82 and 0.67, respectively.           \\ \hline

\textit{NEEDLE} &  This work  &   CNN+DNN      &   ZTF Bright transient survey, \textit{Sherlock}-predicted hosts and Pan-STARRS catalogs  &  Science and reference image in a single band, simple light curve and most galaxy metadata.     &  For SLSNe-I, averaged completeness is 0.77, averaged purity is 0.82 in the test sets. For TDE, the numbers are 0.72 and 0.79.   \\ \hline
\end{tabular}
\end{adjustbox}
\caption{Comparisons among various transient classifiers for SLSNe and TDEs.}
\label{tab: comparison with classifiers}
\end{table*}


\subsection{Remaining difficulties and future improvement}

While the \textit{NEEDLE} algorithm is performing well on the ZTF data set, we are continuing to develop the code and plan a number of future improvements to deal with current limitations, including:
\begin{itemize}
    \item Unbalanced classes. The rare transients we focus on, including SLSNe and TDE, have less than 100 samples for each. After being split into training and test sets, fewer samples are actually used for model training. Weighted loss functions can solve this problem to a certain extent, but feature extraction of rare classes requires more samples and smarter algorithms, such as small-sample learning.
    \item $NaN$ value replacement and padding. Replacing $NaN$ values with zero poses difficulties for classification, since zero has physical meaning for magnitudes and image pixels. However, given the input requirements of neural networks, some kind of padding is inevitable. The possible solution is to fill in the missing values based on context and modelling.
    \item There is a large fraction of SLSNe without cataloged hosts, and in the early years of LSST, this fraction will increase at higher redshifts, and affect the other classes too. To mitigate this, we will continue to develop and apply the \textit{NEEDLE-T} version of the code in parallel for such cases.
    \item Including more contaminants in our training set. Currently we assume that contaminants such as AGN and variable stars can be rejected by simple \textit{Lasair} filters before they reach \textit{NEEDLE}. This may not be the case in future, deeper surveys like LSST.
    \item LSST alert cutouts in real-time are much smaller than for ZTF, and full-size images will only be available after 80 hours. To achieve real-time prediction, older images might need to be included in classification and training, rather than just the most recent detection.
\end{itemize}

Additionally, we have further plans for new features, and analyses to improve our training process. These include:

\begin{itemize}
    \item A detailed study of mis-classified objects. The next step will be to visualize the model behaviour for such objects individually, and try to understand the reasons for mis-classification.
    \item Including more time-domain information. Rather than one image and a set of simple light curve features, using more advanced features, including the light curve directly, or even providing a time series of images, may help to improve performance. For the next Classifier, Conv3D and other relevant networks, such as Recurrent Neural Networks, will be considered.
    \item Early stage classification. The ultimate goal of our classifier is to identify rare events in their early stages, even before their peak. With the addition of images at multiple epochs, we will analyze the trends in accuracy as more observations are added.
\end{itemize}

\section{Conclusion}\label{sec: conclusion}

This paper introduces a novel context-based hybrid neural network, capable of providing probabilistic classifications of transients as SLSNe, TDEs or SNe, at early stages in their evolution.
The literature suggests that SLSNe are typically found in faint, star-forming dwarf galaxies, and TDEs are located at the center of the host galaxies that are often green and centrally concentrated.
Based on the understanding of their unique characteristics, the \textit{NEEDLE} classifier is specifically developed to exploit this information and identify these sources using only single \textit{science} and \textit{reference} image stamps of a transient and its environment, as well as simple photometric information from ZTF (and in future LSST) alert packets, and cataloged host galaxy magnitudes. 

Since half of the hosts of SLSNe are not cataloged, two versions of \textit{NEEDLE} are developed, differentiated according to whether they contain host information. 
Results show that even without a cataloged host galaxy, we are able to identify 79\% of SNe, 76\% SLSNe and 62\% TDEs, averaged among 10 test sets. 
As host information is added, the true positive rate of TDEs increases to approximately 72\%, and the highest true positive rate of SLSNe increases from 87\% to 93\%. To mitigate the issue of contamination from common SNe, we recommend a threshold probability before assigning a classification of $p\gtrsim0.75$. Under these conditions, we can achieve over 95\% completeness for SLSNe and TDEs (on the full data set), at the cost of around 20\% purity.

Furthermore, photometric information has a greater impact on the predictions of SLSN and TDE compared with ordinary SNe, in particular because of their longer rise times. Experiments have shown that the fraction of SLSNe and TDEs classified correctly increases as they rise towards the light curve peak.

Currently, \textit{NEEDLE} is being implemented on the \textit{Lasair} alert broker, and is able to process ZTF streaming alerts and submit an annotation containing the probabilities for three classes back to the broker. These public classifications will help to inform spectroscopic follow-up for these rare events. We are continuing to develop \textit{NEEDLE}, and expect that image-based classification of transients will be a powerful tool in the era of LSST.

\section*{Acknowledgements}

We thank members of the QUB transients discovery team, the QUB Virtual Institute for Data Intensive Research, and the Turing Institute for many helpful conversations. In particular, we thank Aleksandar Novakovic, Richard Gault and Miguel Arana for their advice on neural networks. We also thank Sean McGee and Sebastian Gomez for helpful feedback on the project.

MN and XS are supported by the European Research Council (ERC) under the European Union’s Horizon 2020 research and innovation programme (grant agreement No.~948381). MN also acknowledges support from an Alan Turing Fellowship and UK Space Agency Grant No.~ST/Y000692/1. Lasair is supported by the UKRI Science and Technology Facilities Council and is a collaboration between the University of Edinburgh (grant ST/N002512/1) and Queen’s University Belfast (grant ST/N002520/1) within the LSST:UK Science Consortium.

\section*{Data Availability}

This paper is based on publicly available data. We are making all results from this work publicly available via the LSST:UK \textit{Lasair} broker, and the data repository including the trained models, scripts and HDF5 format data will be made available via \href{https://github.com/XinyueSheng2019/NEEDLE}{Github}.



\bibliographystyle{mnras}
\bibliography{main} 

\begin{thebibliography}{}
\makeatletter
\relax
\def\mn@urlcharsother{\let\do\@makeother \do\$\do\&\do\#\do\^\do\_\do\%\do\~}
\def\mn@doi{\begingroup\mn@urlcharsother \@ifnextchar [ {\mn@doi@}
  {\mn@doi@[]}}
\def\mn@doi@[#1]#2{\def\@tempa{#1}\ifx\@tempa\@empty \href
  {http://dx.doi.org/#2} {doi:#2}\else \href {http://dx.doi.org/#2} {#1}\fi
  \endgroup}
\def\mn@eprint#1#2{\mn@eprint@#1:#2::\@nil}
\def\mn@eprint@arXiv#1{\href {http://arxiv.org/abs/#1} {{\tt arXiv:#1}}}
\def\mn@eprint@dblp#1{\href {http://dblp.uni-trier.de/rec/bibtex/#1.xml}
  {dblp:#1}}
\def\mn@eprint@#1:#2:#3:#4\@nil{\def\@tempa {#1}\def\@tempb {#2}\def\@tempc
  {#3}\ifx \@tempc \@empty \let \@tempc \@tempb \let \@tempb \@tempa \fi \ifx
  \@tempb \@empty \def\@tempb {arXiv}\fi \@ifundefined
  {mn@eprint@\@tempb}{\@tempb:\@tempc}{\expandafter \expandafter \csname
  mn@eprint@\@tempb\endcsname \expandafter{\@tempc}}}

\bibitem[\protect\citeauthoryear{Angus, Levan, Perley, Tanvir, Lyman, Stanway
  \& Fruchter}{Angus et~al.}{2016}]{Angus_2016}
Angus C.~R.,  Levan A.~J.,  Perley D.~A.,  Tanvir N.~R.,  Lyman J.~D.,  Stanway
  E.~R.,   Fruchter A.~S.,  2016, \mn@doi [Monthly Notices of the Royal
  Astronomical Society] {10.1093/mnras/stw063}, 458, 84

\bibitem[\protect\citeauthoryear{{Baldeschi}, {Miller}, {Stroh}, {Margutti}  \&
  {Coppejans}}{{Baldeschi} et~al.}{2020}]{Baldeschi_2020}
{Baldeschi} A.,  {Miller} A.,  {Stroh} M.,  {Margutti} R.,   {Coppejans} D.~L.,
   2020, \mn@doi [\apj] {10.3847/1538-4357/abb1c0}, \href
  {https://ui.adsabs.harvard.edu/abs/2020ApJ...902...60B} {902, 60}

\bibitem[\protect\citeauthoryear{Bellm et~al.,}{Bellm
  et~al.}{2018}]{Bellm_2019}
Bellm E.~C.,  et~al., 2018, \mn@doi [Publications of the Astronomical Society
  of the Pacific] {10.1088/1538-3873/aaecbe}, 131, 018002

\bibitem[\protect\citeauthoryear{{Blanchard}, {Berger}  \& {Fong}}{{Blanchard}
  et~al.}{2016}]{Blanchard_2016}
{Blanchard} P.~K.,  {Berger} E.,   {Fong} W.-f.,  2016, \mn@doi [\apj]
  {10.3847/0004-637X/817/2/144}, \href
  {https://ui.adsabs.harvard.edu/abs/2016ApJ...817..144B} {817, 144}

\bibitem[\protect\citeauthoryear{Boone}{Boone}{2019}]{Boone_2019}
Boone K.,  2019, \mn@doi [The Astronomical Journal] {10.3847/1538-3881/ab5182},
  158, 257

\bibitem[\protect\citeauthoryear{{Botticella} et~al.,}{{Botticella}
  et~al.}{2017}]{Botticella_2017}
{Botticella} M.~T.,  et~al., 2017, \mn@doi [\aap]
  {10.1051/0004-6361/201629432}, \href
  {https://ui.adsabs.harvard.edu/abs/2017A&A...598A..50B} {598, A50}

\bibitem[\protect\citeauthoryear{{Bricman} \& {Gomboc}}{{Bricman} \&
  {Gomboc}}{2020}]{Bricman_2020}
{Bricman} K.,  {Gomboc} A.,  2020, \mn@doi [\apj] {10.3847/1538-4357/ab6989},
  \href {https://ui.adsabs.harvard.edu/abs/2020ApJ...890...73B} {890, 73}

\bibitem[\protect\citeauthoryear{{Burhanudin} \& {Maund}}{{Burhanudin} \&
  {Maund}}{2023}]{Burhanudin_2023}
{Burhanudin} U.~F.,  {Maund} J.~R.,  2023, \mn@doi [\mnras]
  {10.1093/mnras/stac3672}, \href
  {https://ui.adsabs.harvard.edu/abs/2023MNRAS.521.1601B} {521, 1601}

\bibitem[\protect\citeauthoryear{{Burhanudin} et~al.,}{{Burhanudin}
  et~al.}{2021}]{Burhanudin_2021}
{Burhanudin} U.~F.,  et~al., 2021, \mn@doi [\mnras] {10.1093/mnras/stab1545},
  \href {https://ui.adsabs.harvard.edu/abs/2021MNRAS.505.4345B} {505, 4345}

\bibitem[\protect\citeauthoryear{Carrasco-Davis et~al.,}{Carrasco-Davis
  et~al.}{2021}]{Carrasco_Davis_2021}
Carrasco-Davis R.,  et~al., 2021, \mn@doi [The Astronomical Journal]
  {10.3847/1538-3881/ac0ef1}, 162, 231

\bibitem[\protect\citeauthoryear{Chambers et~al.,}{Chambers
  et~al.}{2019}]{chambers2019panstarrs1}
Chambers K.~C.,  et~al., 2019, The Pan-STARRS1 Surveys (\mn@eprint {arXiv}
  {1612.05560})

\bibitem[\protect\citeauthoryear{Chen, Smartt, Yates, Nicholl, Krühler,
  Schady, Dennefeld  \& Inserra}{Chen et~al.}{2017}]{Chen_2017}
Chen T.-W.,  Smartt S.~J.,  Yates R.~M.,  Nicholl M.,  Krühler T.,  Schady P.,
   Dennefeld M.,   Inserra C.,  2017, \mn@doi [Monthly Notices of the Royal
  Astronomical Society] {10.1093/mnras/stx1428}, 470, 3566

\bibitem[\protect\citeauthoryear{{Chen} et~al.,}{{Chen}
  et~al.}{2022}]{Chen_2022}
{Chen} Z.~H.,  et~al., 2022, arXiv e-prints, \href
  {https://ui.adsabs.harvard.edu/abs/2022arXiv220202059C} {p. arXiv:2202.02059}

\bibitem[\protect\citeauthoryear{{Cleland}, {McGee}  \& {Nicholl}}{{Cleland}
  et~al.}{2023}]{Cleland_2023}
{Cleland} C.,  {McGee} S.~L.,   {Nicholl} M.,  2023, \mn@doi [\mnras]
  {10.1093/mnras/stad2118}, \href
  {https://ui.adsabs.harvard.edu/abs/2023MNRAS.524.3559C} {524, 3559}

\bibitem[\protect\citeauthoryear{Collette}{Collette}{2013}]{collette_python_hdf5_2014}
Collette A.,  2013, Python and HDF5.
O'Reilly

\bibitem[\protect\citeauthoryear{{Donoso-Oliva}, {Becker}, {Protopapas},
  {Cabrera-Vives}, {Vishnu}  \& {Vardhan}}{{Donoso-Oliva}
  et~al.}{2022}]{Donoso-Oliva_2022}
{Donoso-Oliva} C.,  {Becker} I.,  {Protopapas} P.,  {Cabrera-Vives} G.,
  {Vishnu} M.,   {Vardhan} H.,  2022, arXiv e-prints, \href
  {https://ui.adsabs.harvard.edu/abs/2022arXiv220501677D} {p. arXiv:2205.01677}

\bibitem[\protect\citeauthoryear{{Flewelling} et~al.,}{{Flewelling}
  et~al.}{2020}]{Flewelling_2020}
{Flewelling} H.~A.,  et~al., 2020, \mn@doi [\apjs] {10.3847/1538-4365/abb82d},
  \href {https://ui.adsabs.harvard.edu/abs/2020ApJS..251....7F} {251, 7}

\bibitem[\protect\citeauthoryear{{Foley} \& {Mandel}}{{Foley} \&
  {Mandel}}{2013}]{Foley_2013}
{Foley} R.~J.,  {Mandel} K.,  2013, \mn@doi [\apj]
  {10.1088/0004-637X/778/2/167}, \href
  {https://ui.adsabs.harvard.edu/abs/2013ApJ...778..167F} {778, 167}

\bibitem[\protect\citeauthoryear{{F{\"o}rster} et~al.,}{{F{\"o}rster}
  et~al.}{2022}]{Forster_2022}
{F{\"o}rster} F.,  et~al., 2022, arXiv e-prints, \href
  {https://ui.adsabs.harvard.edu/abs/2022arXiv220804310F} {p. arXiv:2208.04310}

\bibitem[\protect\citeauthoryear{{Fremling} et~al.,}{{Fremling}
  et~al.}{2020}]{Fremling_2020}
{Fremling} C.,  et~al., 2020, \mn@doi [\apj] {10.3847/1538-4357/ab8943}, \href
  {https://ui.adsabs.harvard.edu/abs/2020ApJ...895...32F} {895, 32}

\bibitem[\protect\citeauthoryear{{French}, {Arcavi}  \& {Zabludoff}}{{French}
  et~al.}{2016}]{French_2016}
{French} K.~D.,  {Arcavi} I.,   {Zabludoff} A.,  2016, \mn@doi [\apjl]
  {10.3847/2041-8205/818/1/L21}, \href
  {https://ui.adsabs.harvard.edu/abs/2016ApJ...818L..21F} {818, L21}

\bibitem[\protect\citeauthoryear{{Fruchter} et~al.,}{{Fruchter}
  et~al.}{2006}]{Fruchter_2006}
{Fruchter} A.~S.,  et~al., 2006, \mn@doi [\nat] {10.1038/nature04787}, \href
  {https://ui.adsabs.harvard.edu/abs/2006Natur.441..463F} {441, 463}

\bibitem[\protect\citeauthoryear{{Gagliano}, {Narayan}, {Engel}, {Carrasco
  Kind}  \& {LSST Dark Energy Science Collaboration}}{{Gagliano}
  et~al.}{2021}]{Gagliano_2021}
{Gagliano} A.,  {Narayan} G.,  {Engel} A.,  {Carrasco Kind} M.,   {LSST Dark
  Energy Science Collaboration} 2021, \mn@doi [\apj]
  {10.3847/1538-4357/abd02b}, \href
  {https://ui.adsabs.harvard.edu/abs/2021ApJ...908..170G} {908, 170}

\bibitem[\protect\citeauthoryear{{Gagliano}, {Contardo}, {Foreman Mackey},
  {Malz}  \& {Aleo}}{{Gagliano} et~al.}{2023}]{Gagliano_2023}
{Gagliano} A.,  {Contardo} G.,  {Foreman Mackey} D.,  {Malz} A.~I.,   {Aleo}
  P.~D.,  2023, \mn@doi [arXiv e-prints] {10.48550/arXiv.2305.08894}, \href
  {https://ui.adsabs.harvard.edu/abs/2023arXiv230508894G} {p. arXiv:2305.08894}

\bibitem[\protect\citeauthoryear{{Gal-Yam}}{{Gal-Yam}}{2019}]{Gal-Yam_2019}
{Gal-Yam} A.,  2019, \mn@doi [\araa] {10.1146/annurev-astro-081817-051819},
  \href {https://ui.adsabs.harvard.edu/abs/2019ARA&A..57..305G} {57, 305}

\bibitem[\protect\citeauthoryear{{Gezari}}{{Gezari}}{2021}]{Gezari_2021}
{Gezari} S.,  2021, \mn@doi [\araa] {10.1146/annurev-astro-111720-030029},
  \href {https://ui.adsabs.harvard.edu/abs/2021ARA&A..59...21G} {59, 21}

\bibitem[\protect\citeauthoryear{{Gomez}, {Berger}, {Blanchard},
  {Hosseinzadeh}, {Nicholl}, {Villar}  \& {Yin}}{{Gomez}
  et~al.}{2020}]{Gomez_2020}
{Gomez} S.,  {Berger} E.,  {Blanchard} P.~K.,  {Hosseinzadeh} G.,  {Nicholl}
  M.,  {Villar} V.~A.,   {Yin} Y.,  2020, \mn@doi [\apj]
  {10.3847/1538-4357/abbf49}, \href
  {https://ui.adsabs.harvard.edu/abs/2020ApJ...904...74G} {904, 74}

\bibitem[\protect\citeauthoryear{{Gomez}, {Villar}, {Berger}, {Gezari}, {van
  Velzen}, {Nicholl}, {Blanchard}  \& {Alexander}}{{Gomez}
  et~al.}{2023}]{Gomez_2023}
{Gomez} S.,  {Villar} V.~A.,  {Berger} E.,  {Gezari} S.,  {van Velzen} S.,
  {Nicholl} M.,  {Blanchard} P.~K.,   {Alexander} K.~D.,  2023, \mn@doi [\apj]
  {10.3847/1538-4357/acc535}, \href
  {https://ui.adsabs.harvard.edu/abs/2023ApJ...949..113G} {949, 113}

\bibitem[\protect\citeauthoryear{{Graur}, {Bianco}, {Modjaz}, {Shivvers},
  {Filippenko}, {Li}  \& {Smith}}{{Graur} et~al.}{2017}]{Graur_2017}
{Graur} O.,  {Bianco} F.~B.,  {Modjaz} M.,  {Shivvers} I.,  {Filippenko} A.~V.,
   {Li} W.,   {Smith} N.,  2017, \mn@doi [\apj] {10.3847/1538-4357/aa5eb7},
  \href {https://ui.adsabs.harvard.edu/abs/2017ApJ...837..121G} {837, 121}

\bibitem[\protect\citeauthoryear{{Graur}, {French}, {Zahid}, {Guillochon},
  {Mandel}, {Auchettl}  \& {Zabludoff}}{{Graur} et~al.}{2018}]{Graur_2018}
{Graur} O.,  {French} K.~D.,  {Zahid} H.~J.,  {Guillochon} J.,  {Mandel} K.~S.,
   {Auchettl} K.,   {Zabludoff} A.~I.,  2018, \mn@doi [\apj]
  {10.3847/1538-4357/aaa3fd}, \href
  {https://ui.adsabs.harvard.edu/abs/2018ApJ...853...39G} {853, 39}

\bibitem[\protect\citeauthoryear{Guillochon, Parrent, Kelley  \&
  Margutti}{Guillochon et~al.}{2017}]{Guillochon_2017}
Guillochon J.,  Parrent J.,  Kelley L.~Z.,   Margutti R.,  2017, \mn@doi [The
  Astrophysical Journal] {10.3847/1538-4357/835/1/64}, 835, 64

\bibitem[\protect\citeauthoryear{{Hammerstein} et~al.,}{{Hammerstein}
  et~al.}{2022}]{Hammerstein_2022}
{Hammerstein} E.,  et~al., 2022, arXiv e-prints, \href
  {https://ui.adsabs.harvard.edu/abs/2022arXiv220301461H} {p. arXiv:2203.01461}

\bibitem[\protect\citeauthoryear{{Hills}}{{Hills}}{1975}]{Hills_1975}
{Hills} J.~G.,  1975, \mn@doi [\nat] {10.1038/254295a0}, \href
  {https://ui.adsabs.harvard.edu/abs/1975Natur.254..295H} {254, 295}

\bibitem[\protect\citeauthoryear{{Hlo{\v{z}}ek} et~al.,}{{Hlo{\v{z}}ek}
  et~al.}{2023}]{Hlovzek_2023}
{Hlo{\v{z}}ek} R.,  et~al., 2023, \mn@doi [\apjs] {10.3847/1538-4365/accd6a},
  \href {https://ui.adsabs.harvard.edu/abs/2023ApJS..267...25H} {267, 25}

\bibitem[\protect\citeauthoryear{{Hosseinzadeh} et~al.,}{{Hosseinzadeh}
  et~al.}{2020}]{Hosseinzadeh_2020}
{Hosseinzadeh} G.,  et~al., 2020, \mn@doi [\apj] {10.3847/1538-4357/abc42b},
  \href {https://ui.adsabs.harvard.edu/abs/2020ApJ...905...93H} {905, 93}

\bibitem[\protect\citeauthoryear{{Hsu}, {Tan}, {Holdship}, {Duo}, {Xu}, {Viti},
  {Wu}  \& {Gaches}}{{Hsu} et~al.}{2023}]{Hsu_2023}
{Hsu} C.-J.,  {Tan} J.~C.,  {Holdship} J.,  {Duo} {Xu} {Viti} S.,  {Wu} B.,
  {Gaches} B.,  2023, \mn@doi [arXiv e-prints] {10.48550/arXiv.2308.11803},
  \href {https://ui.adsabs.harvard.edu/abs/2023arXiv230811803H} {p.
  arXiv:2308.11803}

\bibitem[\protect\citeauthoryear{{Ivezi{\'c}} et~al.,}{{Ivezi{\'c}}
  et~al.}{2019}]{Ivezic_2019}
{Ivezi{\'c}} {\v{Z}}.,  et~al., 2019, \mn@doi [\apj]
  {10.3847/1538-4357/ab042c}, \href
  {https://ui.adsabs.harvard.edu/abs/2019ApJ...873..111I} {873, 111}

\bibitem[\protect\citeauthoryear{{Kantor}}{{Kantor}}{2014}]{Kantor_2014}
{Kantor} J.,  2014, in {Wozniak} P.~R.,  {Graham} M.~J.,  {Mahabal} A.~A.,
  {Seaman} R.,  eds, The Third Hot-wiring the Transient Universe Workshop. pp
  19--26

\bibitem[\protect\citeauthoryear{{Kelly} \& {Kirshner}}{{Kelly} \&
  {Kirshner}}{2012}]{Kelly_2012}
{Kelly} P.~L.,  {Kirshner} R.~P.,  2012, \mn@doi [\apj]
  {10.1088/0004-637X/759/2/107}, \href
  {https://ui.adsabs.harvard.edu/abs/2012ApJ...759..107K} {759, 107}

\bibitem[\protect\citeauthoryear{{Kessler} et~al.,}{{Kessler}
  et~al.}{2019}]{Kessler_2019}
{Kessler} R.,  et~al., 2019, \mn@doi [\pasp] {10.1088/1538-3873/ab26f1}, \href
  {https://ui.adsabs.harvard.edu/abs/2019PASP..131i4501K} {131, 094501}

\bibitem[\protect\citeauthoryear{{Kingma} \& {Ba}}{{Kingma} \&
  {Ba}}{2014}]{Kingma_2014}
{Kingma} D.~P.,  {Ba} J.,  2014, arXiv e-prints, \href
  {https://ui.adsabs.harvard.edu/abs/2014arXiv1412.6980K} {p. arXiv:1412.6980}

\bibitem[\protect\citeauthoryear{Kisley, Qin, Zabludoff, Barnard  \& Ko}{Kisley
  et~al.}{2022}]{Kisley_2022}
Kisley M.,  Qin Y.-J.,  Zabludoff A.,  Barnard K.,   Ko C.-L.,  2022,
  Classifying Astronomical Transients Using Only Host Galaxy Photometry,
  \mn@doi{10.48550/ARXIV.2209.02784}, \url {https://arxiv.org/abs/2209.02784}

\bibitem[\protect\citeauthoryear{{Law-Smith}, {Ramirez-Ruiz}, {Ellison}  \&
  {Foley}}{{Law-Smith} et~al.}{2017}]{Law-Smith_2017}
{Law-Smith} J.,  {Ramirez-Ruiz} E.,  {Ellison} S.~L.,   {Foley} R.~J.,  2017,
  \mn@doi [\apj] {10.3847/1538-4357/aa94c7}, \href
  {https://ui.adsabs.harvard.edu/abs/2017ApJ...850...22L} {850, 22}

\bibitem[\protect\citeauthoryear{Leloudas et~al.,}{Leloudas
  et~al.}{2015}]{Leloudas_2015}
Leloudas G.,  et~al., 2015, \mn@doi [Monthly Notices of the Royal Astronomical
  Society] {10.1093/mnras/stv320}, 449, 917

\bibitem[\protect\citeauthoryear{{Li}, {Chornock}, {Leaman}, {Filippenko},
  {Poznanski}, {Wang}, {Ganeshalingam}  \& {Mannucci}}{{Li}
  et~al.}{2011}]{Li_2011}
{Li} W.,  {Chornock} R.,  {Leaman} J.,  {Filippenko} A.~V.,  {Poznanski} D.,
  {Wang} X.,  {Ganeshalingam} M.,   {Mannucci} F.,  2011, \mn@doi [\mnras]
  {10.1111/j.1365-2966.2011.18162.x}, \href
  {https://ui.adsabs.harvard.edu/abs/2011MNRAS.412.1473L} {412, 1473}

\bibitem[\protect\citeauthoryear{{Li} et~al.,}{{Li} et~al.}{2022}]{Li_2022}
{Li} R.,  et~al., 2022, arXiv e-prints, \href
  {https://ui.adsabs.harvard.edu/abs/2022arXiv220510720L} {p. arXiv:2205.10720}

\bibitem[\protect\citeauthoryear{{Lunnan} et~al.,}{{Lunnan}
  et~al.}{2014}]{Lunnan_2014}
{Lunnan} R.,  et~al., 2014, \mn@doi [\apj] {10.1088/0004-637X/787/2/138}, \href
  {https://ui.adsabs.harvard.edu/abs/2014ApJ...787..138L} {787, 138}

\bibitem[\protect\citeauthoryear{{Miranda} et~al.,}{{Miranda}
  et~al.}{2022}]{Miranda_2022}
{Miranda} N.,  et~al., 2022, arXiv e-prints, \href
  {https://ui.adsabs.harvard.edu/abs/2022arXiv220806534M} {p. arXiv:2208.06534}

\bibitem[\protect\citeauthoryear{Muthukrishna, Narayan, Mandel, Biswas  \&
  Hlo{\v{z}}ek}{Muthukrishna et~al.}{2019}]{Muthukrishna_2019}
Muthukrishna D.,  Narayan G.,  Mandel K.~S.,  Biswas R.,   Hlo{\v{z}}ek R.,
  2019, \mn@doi [Publications of the Astronomical Society of the Pacific]
  {10.1088/1538-3873/ab1609}, 131, 118002

\bibitem[\protect\citeauthoryear{{Nicholl}}{{Nicholl}}{2021}]{Nicholl_2021}
{Nicholl} M.,  2021, \mn@doi [Astronomy and Geophysics]
  {10.1093/astrogeo/atab092}, \href
  {https://ui.adsabs.harvard.edu/abs/2021A&G....62.5.34N} {62, 5.34}

\bibitem[\protect\citeauthoryear{O'Malley, Bursztein, Long, Chollet, Jin,
  Invernizzi  et~al.}{O'Malley et~al.}{2019}]{omalley_2019}
O'Malley T.,  Bursztein E.,  Long J.,  Chollet F.,  Jin H.,  Invernizzi L.,
  et~al., 2019, KerasTuner, \url{https://github.com/keras-team/keras-tuner}

\bibitem[\protect\citeauthoryear{{{\O}rum}, {Ivens}, {Strandberg}, {Leloudas},
  {Man}  \& {Schulze}}{{{\O}rum} et~al.}{2020}]{Orum_2020}
{{\O}rum} S.~V.,  {Ivens} D.~L.,  {Strandberg} P.,  {Leloudas} G.,  {Man} A.
  W.~S.,   {Schulze} S.,  2020, \mn@doi [\aap] {10.1051/0004-6361/202038176},
  \href {https://ui.adsabs.harvard.edu/abs/2020A&A...643A..47O} {643, A47}

\bibitem[\protect\citeauthoryear{{Perley} et~al.,}{{Perley}
  et~al.}{2016}]{Perley_2016}
{Perley} D.~A.,  et~al., 2016, \mn@doi [\apj] {10.3847/0004-637X/830/1/13},
  \href {https://ui.adsabs.harvard.edu/abs/2016ApJ...830...13P} {830, 13}

\bibitem[\protect\citeauthoryear{{Perley} et~al.,}{{Perley}
  et~al.}{2020}]{Perley_2020}
{Perley} D.~A.,  et~al., 2020, \mn@doi [\apj] {10.3847/1538-4357/abbd98}, \href
  {https://ui.adsabs.harvard.edu/abs/2020ApJ...904...35P} {904, 35}

\bibitem[\protect\citeauthoryear{Pimentel, Est{\'{e}}vez  \& Förster}{Pimentel
  et~al.}{2022}]{Pimentel_2022}
Pimentel {\'{O} }.,  Est{\'{e}}vez P.~A.,   Förster F.,  2022, \mn@doi [The
  Astronomical Journal] {10.3847/1538-3881/ac9ab4}, 165, 18

\bibitem[\protect\citeauthoryear{{Qu} \& {Sako}}{{Qu} \&
  {Sako}}{2022}]{Qu_2022}
{Qu} H.,  {Sako} M.,  2022, \mn@doi [\aj] {10.3847/1538-3881/ac39a1}, \href
  {https://ui.adsabs.harvard.edu/abs/2022AJ....163...57Q} {163, 57}

\bibitem[\protect\citeauthoryear{{Quimby} et~al.,}{{Quimby}
  et~al.}{2011}]{Quimby_2011}
{Quimby} R.~M.,  et~al., 2011, \mn@doi [\nat] {10.1038/nature10095}, \href
  {https://ui.adsabs.harvard.edu/abs/2011Natur.474..487Q} {474, 487}

\bibitem[\protect\citeauthoryear{{Ramsden}, {Lanning}, {Nicholl}  \&
  {McGee}}{{Ramsden} et~al.}{2022}]{Ramsden_2022}
{Ramsden} P.,  {Lanning} D.,  {Nicholl} M.,   {McGee} S.~L.,  2022, \mn@doi
  [\mnras] {10.1093/mnras/stac1810}, \href
  {https://ui.adsabs.harvard.edu/abs/2022MNRAS.515.1146R} {515, 1146}

\bibitem[\protect\citeauthoryear{{Rees}}{{Rees}}{1988}]{Rees_1988}
{Rees} M.~J.,  1988, \mn@doi [\nat] {10.1038/333523a0}, \href
  {https://ui.adsabs.harvard.edu/abs/1988Natur.333..523R} {333, 523}

\bibitem[\protect\citeauthoryear{{S{\'a}nchez-S{\'a}ez}
  et~al.,}{{S{\'a}nchez-S{\'a}ez} et~al.}{2021}]{Sanchez-Saez_2021}
{S{\'a}nchez-S{\'a}ez} P.,  et~al., 2021, \mn@doi [\aj]
  {10.3847/1538-3881/abd5c1}, \href
  {https://ui.adsabs.harvard.edu/abs/2021AJ....161..141S} {161, 141}

\bibitem[\protect\citeauthoryear{Schulze et~al.,}{Schulze
  et~al.}{2017}]{Schulze_2017}
Schulze S.,  et~al., 2017, \mn@doi [Monthly Notices of the Royal Astronomical
  Society] {10.1093/mnras/stx2352}, 473, 1258

\bibitem[\protect\citeauthoryear{{Shappee} et~al.,}{{Shappee}
  et~al.}{2014}]{Shappee_2014}
{Shappee} B.,  et~al., 2014, in American Astronomical Society Meeting Abstracts
  \#223. p. 236.03

\bibitem[\protect\citeauthoryear{{Smith} et~al.,}{{Smith}
  et~al.}{2019}]{Smith_2019}
{Smith} K.~W.,  et~al., 2019, \mn@doi [Research Notes of the American
  Astronomical Society] {10.3847/2515-5172/ab020f}, \href
  {https://ui.adsabs.harvard.edu/abs/2019RNAAS...3...26S} {3, 26}

\bibitem[\protect\citeauthoryear{{Smith} et~al.,}{{Smith}
  et~al.}{2020}]{Smith_2020}
{Smith} K.~W.,  et~al., 2020, \mn@doi [\pasp] {10.1088/1538-3873/ab936e}, \href
  {https://ui.adsabs.harvard.edu/abs/2020PASP..132h5002S} {132, 085002}

\bibitem[\protect\citeauthoryear{{Stein} et~al.,}{{Stein}
  et~al.}{2023}]{Stein_2023}
{Stein} R.,  et~al., 2023, arXiv e-prints, \href
  {https://ui.adsabs.harvard.edu/abs/2023arXiv231200139S} {p. arXiv:2312.00139}

\bibitem[\protect\citeauthoryear{{Sullivan} et~al.,}{{Sullivan}
  et~al.}{2006}]{Sullivan_2006}
{Sullivan} M.,  et~al., 2006, \mn@doi [\apj] {10.1086/506137}, \href
  {https://ui.adsabs.harvard.edu/abs/2006ApJ...648..868S} {648, 868}

\bibitem[\protect\citeauthoryear{{Tonry} et~al.,}{{Tonry}
  et~al.}{2018}]{Tonry_2018}
{Tonry} J.~L.,  et~al., 2018, \mn@doi [\pasp] {10.1088/1538-3873/aabadf}, \href
  {https://ui.adsabs.harvard.edu/abs/2018PASP..130f4505T} {130, 064505}

\bibitem[\protect\citeauthoryear{Villar, Nicholl  \& Berger}{Villar
  et~al.}{2018}]{Villar_2018}
Villar V.~A.,  Nicholl M.,   Berger E.,  2018, \mn@doi [The Astrophysical
  Journal] {10.3847/1538-4357/aaee6a}, 869, 166

\bibitem[\protect\citeauthoryear{{Villar} et~al.,}{{Villar}
  et~al.}{2020}]{Villar_2020}
{Villar} V.~A.,  et~al., 2020, \mn@doi [\apj] {10.3847/1538-4357/abc6fd}, \href
  {https://ui.adsabs.harvard.edu/abs/2020ApJ...905...94V} {905, 94}

\bibitem[\protect\citeauthoryear{{Yao} et~al.,}{{Yao} et~al.}{2023}]{Yao_2023}
{Yao} Y.,  et~al., 2023, \mn@doi [\apjl] {10.3847/2041-8213/acf216}, \href
  {https://ui.adsabs.harvard.edu/abs/2023ApJ...955L...6Y} {955, L6}

\bibitem[\protect\citeauthoryear{{van Velzen} et~al.,}{{van Velzen}
  et~al.}{2021}]{van_Velzen_2021}
{van Velzen} S.,  et~al., 2021, \mn@doi [\apj] {10.3847/1538-4357/abc258},
  \href {https://ui.adsabs.harvard.edu/abs/2021ApJ...908....4V} {908, 4}

\makeatother
\end{thebibliography}




\appendix

\section{ZTF objects type}

\begin{table*}
\begin{tabular}{|c|c|p{200pt}|c|}
\hline
Label & Type & Feature & Number \\ \hline
\multirow{3}{*}{\begin{tabular}[c]{@{}c@{}}SN Ia\\ (4113)\end{tabular}} & Ia & Thermonuclear explosion of white dwarf; spectrum lacks hydrogen and helium & 4095 \\ \cline{2-4} 
& Iax & A faint and fast sub-class of SNe Ia & 11            \\ \cline{2-4} 
 & Ia-CSM & SN Ia interacting with nearby circumstellar material & 7 \\ \hline

\begin{tabular}[c]{@{}c@{}} SN II\\ (899)\end{tabular} & II & Core-collapse explosion of a red supergiant $\gtrsim 8$\,M$_\odot$ & 899 \\ \hline

\multirow{4}{*}{\begin{tabular}[c]{@{}c@{}}Stripped Envelope SN\\ (363)\end{tabular}} & Ib/c & Massive stars that have lost their hydrogen (Ib) or hydrogen and helium layers (Ic)  & 216 \\ \cline{2-4} 
& IIb & Incomplete envelope stripping; initially show hydrogen lines, but quickly change to resemble a SN Ib & 80 \\ \cline{2-4} 
 & Ic-BL & Broad spectral lines due to high velocities, large nickel masses, the only SN type associated with gamma-ray bursts & 47 \\ \cline{2-4} 
 & Ibn & Supernova interacting with a helium-rich CSM  & 20 \\ \\ \hline

\multirow{2}{*}{\begin{tabular}[c]{@{}c@{}}Interacting SN\\ (211)\end{tabular}} & IIn & Hydrogen emission lines with narrow Doppler widths, indicating low-velocity CSM that has been shock-excited by a collision from the supernova ejecta & 183 \\  \cline{2-4} 
&SLSN II & IIn brighter than -21 mag & 28 \\\hline

\begin{tabular}[c]{@{}c@{}}SLSN\\ (87)\end{tabular} & SLSN & 10-100 times brighter than normal SN, no Hydrogen and usually no helium; late spectra resemble SNe Ic. Prefer dwarf galaxies.  & 87 \\ \hline

\begin{tabular}[c]{@{}c@{}}TDE\\ (64)\end{tabular} & TDE & A star approaches close to a supermassive black hole and is pulled apart by tidal forces, leading to fallback and accretion & 64 \\ \hline
\multirow{3}{*}{\begin{tabular}[c]{@{}c@{}}Other\\ (18)\end{tabular}} & Gap & Transient with luminosity intermediate between typical SNe and classical novae & 13 \\ \cline{2-4} 
& Ca-rich & Faint and fast transients of ambiguous nature, with strong calcium lines in spectrum & 3 \\ \cline{2-4} 
 & Other &  & 2 \\ \hline
\multirow{4}{*}{\begin{tabular}[c]{@{}c@{}}Non-SN\\ (40)\end{tabular}} & Novae & Outburst on the surface of an accreting white dwarf & 30 \\ \cline{2-4} 
 & ILRT & Intermediate Luminosity Red Transient & 4 \\ \cline{2-4} 
 & LBV & Luminous Blue Variable: very massive star undergoing eruptive mass loss & 4 \\ \cline{2-4}
  & LRN & Luminous Red Novae: mergers of low-mass stellar binaries & 2 \\  \hline
Sum &  &  & 5795 \\ \hline
\end{tabular}
\caption{Summary of ZTF transients with their types, features and numbers. Our ``SN'' class includes the first four labels in this table, but in future versions we aim to resolve these SN sub-types}
\label{tab: all_objects}
\end{table*}

\end{document}